\documentclass[12pt]{article}

\pdfoutput=1

\usepackage{ifpdf}
\ifpdf
\usepackage{graphicx,color}
\usepackage{hyperref}
\else
\usepackage[dvipdfmx]{graphicx,color}
\usepackage[dvipdfmx]{hyperref}
\fi
\usepackage{amssymb, amsfonts, amsmath, cancel, cite, multirow}
\usepackage{tikz}
\usepackage[capitalise]{cleveref}
\usepackage[margin=1in]{geometry}
\usepackage{braket}
\usepackage{ulem}
\newcommand{\eqs}[1]{\begin{equation}\begin{split} #1 \end{split}\end{equation}}

\begin{document}

\begin{titlepage}

\begin{flushright}
\end{flushright}

\vskip 1.35cm
\begin{center}

{\large
\textbf{
    Mediator Decay through mixing with Degenerate Spectrum
}
}
\vskip 1.2cm

\mbox{
	Ayuki Kamada$^{a}$,
	Takumi Kuwahara$^{b}$,
	Shigeki Matsumoto$^{c}$,
        Yu Watanabe$^{c}$,
	and Yuki Watanabe$^{c}$
}

\vskip 0.4cm

\textit{$^a$
    Institute of Theoretical Physics, Faculty of Physics, University of Warsaw, ul. Pasteura 5, PL-02-093 Warsaw, Poland
}

\textit{$^b$
    Center for High Energy Physics, Peking University, Beijing 100871, China
}

\textit{$^c$
    Kavli IPMU (WPI), UTIAS,
    University of Tokyo, Kashiwa, Chiba 277-8583, Japan
}

\vskip 1.5cm
\begin{abstract}
    \noindent
    The decay of the mediator particle into standard model (SM) particles plays a significant role in exploring the dark sector scenario. 
    We consider such a decay, taking the dark photon mediator as an example that mixes with the SM photon. 
    We find that it requires a careful analysis of the decay rate in the presence of an SM vector boson (e.g., $Z$ boson, $\rho$ meson, and true muonium, etc.) nearly degenerate with the mediator particle in mass.
    The decay rate of the mediator particle calculated in the mass eigenstate basis \textbf{does not} agree with the correct result, given by the imaginary parts of the poles for the vector boson propagators, when the mixing parameter is smaller than a specific value.
    In such a case, the decay rate calculated by treating the mixing as a perturbative parameter is in agreement with the correct result. 
    We clarify specific values for the mixing parameter quantitatively using several concrete examples of the SM vector bosons degenerate with the dark photon. 
    When the mass mixing between the vector boson and dark photon is smaller (larger) than the decay width of the vector boson, the latter (former) method to calculate the decay rate of the mediator particle gives the correct result. 
\end{abstract}
\end{center}
\end{titlepage}

\section{Introduction}

Decay is one of the essential features of unstable particles. It plays a crucial role in particle physics for discovering new particles in terrestrial experiments and cosmological observations. Dark-sector models have been extensively studied among various new physics models over the last decade. The dark-sector particles are neutral under the standard model (SM) gauge symmetries, have their own gauge interactions, and are not directly coupled to the SM particles. Some dark-sector particles, referred to as mediator particles in this article, mix with the SM neutral particles, such as the photon, the neutral component of the Higgs boson, and the neutrino. When the mediator particle is lightest in the dark sector, it decays only into the SM particles through the mixing. Its decay provides promising signals to explore the dark sector, e.g., the impact on the cosmic history\,\cite{Fradette:2014sza, Kamada:2015era, Hufnagel:2018bjp, Ibe:2018juk, Ibe:2019gpv}, and the visible (long-lived) decay signal at collider and fixed-target experiments\,\cite{APEX:2011dww, KLOE-2:2011hhj, KLOE-2:2012lii, BaBar:2014zli, LHCb:2015nkv, NA482:2015wmo, LHCb:2016awg, LHCb:2017trq, NA62:2020pwi, NA62:2021zjw, CMS:2021sch, Feng:2022inv, Belle-II:2023ueh, FASER:2023tle}.

The mass of the mediator particle is still unknown, and a wide range of the mass must be explored. In some cases, the mass of the mediator field could be very close to that of an SM neutral field, which is not necessarily the fundamental one but also a resonance or a bound state composed of the SM elementary particles. In this article, we refer to it as a degenerate partner. When the mediator mass is very close to that of the degenerate partner, the mediator field almost maximally mixes with the partner field. In such a case, we na\"ively expect that both the decay widths of these mass eigenstates to the SM particles are comparable. On the other hand, as far as the mixing parameter is considered small, i.e., a perturbative one, the decay width of the mediator particle is expected to be suppressed by the mixing parameter. Hence, the width of the mediator particle becomes smaller than that of the degenerate partner. Since these two statements above seem incompatible, there must be a criterion to specify a parameter region where one statement is valid and the other is not.  

The total decay width of the mediator particle is computed in several ways. 
One way is taking a mass basis, where the mixing among the fields is removed by redefining the fields (e.g., via the diagonalization of their mass matrix) and computing the decay width of the mediator(-like) particles in a standard perturbation manner. In this method, which we refer to as the ``classical'' method in this article, the decay rates of both the mass eigenstates approach each other when they are nearly degenerate in mass.
This classical method is broadly used in the literature; for example, in the electroweak precision measurements on the dark photon that is nearly degenerate with the $Z$ boson~\cite{Hook:2010tw}.
Meanwhile, when the mass mixing between them is treated as a perturbative one, the decay width of the mediator particle is computed by inserting the mass mixing into the decay amplitude: the mediator particle is converted into the degenerate partner by the insertion, and the partner decays into the SM particles. 
This method corresponds to the second statement mentioned above, which we refer to as the ``mass-insertion'' method.
In this method, the mixing parameter suppresses the decay width of the mediator particle and furthermore the decay width of the mediator particle is proportional to the inverse of the decay width of the degenerate partner when the mediator particle and the degenerate partner are nearly degenerate in mass. 
The application of the ``mass-insertion'' method is widely found in the context of the long-lived particle searches, \textit{e.g.}, for the dark photon searches at fixed-target experiments~\cite{Bjorken:2009mm,Blumlein:2013cua,Berlin:2018jbm,Ilten:2018crw}.
The final method, which we call the pole method in this article, is to compute the total decay width using the imaginary part of the propagator pole, corresponding to extracting the pole singularity of some scattering amplitude.
This pole method is also broadly employed in the literature: nearly-degenerate systems with a new vector boson ($Z'$)~\cite{Qiu:2023zfr,LoChiatto:2024guj} and with a new scalar boson (dark Higgs)~\cite{Boyanovsky:2017esz,Sakurai:2022cki}. Assuming this method provides the correct definition of the decay width, we find that the ``classical''\,(``mass-insertion'') method is validated as far as the mass mixing is larger\,(smaller) than the decay width of the degenerate partner. 
We also demonstrate in this article that our criterion works well using several concrete examples, where the validity of the approximation methods\,(the first two methods mentioned above) is quantitatively figured out by comparing with the rigorous method\,(the last method).

We take the dark photon model as a concrete example of the mediator field, which has a kinetic mixing with the SM photon\,\cite{Holdom:1985ag}. The dark photon also mixes with various massive vector fields; we consider the $Z$ boson, $\rho$ meson, and true muonium ($\mu^+ \mu^-$ bound state) as such a vector field (i.e., the degenerate partner), covering a wide range of mass and decay width. Then the dark photon decays into the SM particles by mixing with the SM photon and the massive vector fields.

This paper is organized as follows. In \cref{sec:Med_Decay}, we give a general discussion of the decay rate of the mediator particle by considering the dark photon model as a concrete example. We introduce three methods for computing the total decay rate mentioned above: the ``classical'' method, the ``mass-insertion'' method, and the decay rate given by the imaginary part of the propagator pole. There, we find the criterion for the validity of the ``classical'' and the ``mass-insertion'' methods. In \cref{sec:DP_Decay}, we discuss the decay of the dark photon in the presence of a specific degenerate partner, considering the $Z$ boson, the $\rho$ meson, and the true muonium as examples of such partner particles. In \cref{sec:Dis}, we summarize our criterion for the total decay width of the mediator particle and apply it to other resonances not treated in \cref{sec:DP_Decay}. \cref{sec:Sum} is devoted to concluding our study.

\section{Mediator particle decay}
\label{sec:Med_Decay}

In this section, we calculate the decay rate of a mediator particle that is degenerate with some SM particle in mass. We consider the case with a vector mediator particle, i.e., the so-called dark photon model, which kinematically mixes with the SM photon (or the hypercharge gauge boson). The Lagrangian of this dark photon model is given as follows: 
\eqs{
	\mathcal{L}
	= \mathcal{L}_\mathrm{SM} 
	+ \mathcal{L}_\mathrm{DS} 
	- \frac14 F'_{\mu\nu} F'^{\mu\nu} 
	+ \frac12 \overline m_{A'}^2 A'^2_\mu 
	+ \frac{\epsilon}{2 \cos\theta_W} F'_{\mu\nu} B^{\mu\nu} \,.
	\label{eq:Lagrangian_SM-DS}
}
Here, $A'_\mu$ and $F'_{\mu\nu}$ denote the dark photon field and its field strength tensor, respectively, while $B_{\mu\nu}$ is the field strength tensor of the hypercharge gauge boson, with $\theta_W$ being the Weinberg angle. The first two terms, $\mathcal{L}_\mathrm{SM}$ and $\mathcal{L}_\mathrm{DS}$, are the Lagrangians of the SM sector and the dark sector, respectively. The dark photon couples not only to the SM photon but also to the $Z$ boson after the electroweak symmetry is spontaneously broken. Moreover, the SM photon may mix with other massive SM vector bosons in the low-energy spectrum, such as hadronic vector resonances (e.g., $\rho, \omega, J/\psi$, etc.) and other bound states (e.g., the positronium $e^+ e^-$, the true muonium $\mu^+ \mu^-$, etc.).

We now consider the system containing three vector fields: the SM photon $A_\mu$, the dark photon $A'_\mu$, and a massive SM vector boson (i.e., the degenerate partner) $V_\mu$. In the next section, we discuss several concrete examples of the vector boson $V$, such as the $Z$ boson, the $\rho$ meson, and the true muonium $\mu^+ \mu^-$. While the vector boson $V$ has the kinetic mixing with the SM photon $A$, $V$ may also have kinetic mixing with the dark photon $A'$ from the first beginning. Even if it is not the case, one can obtain such a kinetic mixing when one removes the kinetic mixing term between $V$ and the SM photon $A$ by shifting $A$. The field redefinition of the vector field $V$ then absorbs the kinetic mixing between $V$ and $A'$, and one obtains the mass mixing between the two fields. The mass matrix is generally written as
\eqs{
	M_{V,A'}^2 = \overline m_V^2
	\begin{pmatrix}
		1 & - \eta \\
		-\eta & \eta^2 + \delta^2
	\end{pmatrix} \,,
	\label{eq:massmat}
}
where $\overline m_V^2$ is the mass parameter of $V$ in the interaction basis (with only the kinetic mixing between $V$ and $A'$). The off-diagonal component $\eta$ arises from the kinetic mixing between $A'$ and $V$, and $\delta$ is the mass ratio $\overline m_{A'}/\overline m_V$. We obtain the mass eigenvalues and the mixing angle as
\eqs{
	m_{X\,,Y}^2 & = \frac{\overline{m}_V^2}{2} \left[ 1 + \eta^2 + \delta^2 \pm (1 - \eta^2 - \delta^2) \cos 2 \theta \mp \eta \sin 2\theta \right] \,, \\
	\tan 2\theta & = - \frac{2\eta}{1-\eta^2-\delta^2} \,.
	\label{eq:mass_mixing}
}
Here, the mass mixing angle $\theta$ ranges from $- \pi/4$ to $\pi/4$ and $m_{X\,,Y}^2$ gives the mass eigenvalues of $V$ and $A'$, respectively, except for $\delta$ being close to unity. Since the mass mixing angle $\theta$ jumps from $\mp \pi/4$ to $\pm \pi/4$ (the sign depends on the sign of $\eta$) at $\delta \simeq 1$, the mass eigenvalues have a mass gap of $\eta$. This raises a question of how we identify the particles near $\delta \simeq 1$. As we will see later, the decay rates of the mass eigenstates are identical at a certain $\delta \simeq 1$, and hence, the rates are smooth as a function of $\delta$, unlike the mass eigenvalues. We thus identify the physical states of the particles $V$ and $A'$ by those whose decay rates at $\delta \simeq 1$ are smoothly connected to their decay rates at $\delta$ far from 1. As a result, the relation between the mass eigenstates and the particles $V$ and $A'$ interchange twice around $\delta \simeq 1$.

Suppose $V$ couples to $f$ in the interaction basis as 
\eqs{
	\mathcal{L}_\mathrm{int} 
	& = g_V V_\mu \overline f \gamma^\mu f + e Q_f A_\mu \overline f \gamma^\mu f \,.
}
Here, $g_V$ and $e$ are coupling constants, and $Q_f$ is the electromagnetic charge of the fermion $f$. The mixing matrix relating the interaction basis and the mass basis, where all the kinetic and mass mixing are removed, is obtained as follows: 
\eqs{
	\begin{pmatrix}
		A'_\mu \\
		V_\mu \\
		A_\mu
	\end{pmatrix}_\mathrm{int}
	=
	C_\mathrm{kin}
	\begin{pmatrix}
		\cos\theta & \sin\theta & 0 \\
		- \sin\theta & \cos\theta & 0 \\
		0 & 0 & 1
	\end{pmatrix}
	\begin{pmatrix}
		A'_\mu \\
		V_\mu \\
		A_\mu
	\end{pmatrix}_\mathrm{mass}
	\equiv C \times 
	\begin{pmatrix}
		A'_\mu \\
		V_\mu \\
		A_\mu
	\end{pmatrix}_\mathrm{mass} \,,
	\label{eq:mass_basis}
}
where $C_\mathrm{kin}$ is the mixing matrix for diagonalization (canonicalization) of the kinetic terms and depends on the kinetic mixing between the SM photon and the vector field $V$ (or the hypercharge gauge boson and the weak gauge boson). 

We now introduce three methods to calculate the total decay width of the dark photon $A'$, assuming that $A'$ is the lightest particle in the dark sector and that a single decay channel into a fermion pair $A' \to f \bar{f}$ dominates its width. 
We discuss these three methods in the SM meson systems in \cref{app:hadron}, in particular the $\rho$-$\omega$ mixing in \cref{app:rhoomegagamma} and the kaon physics in \cref{app:kaon}.

\subsubsection*{``Classical'' method}
We calculate the width in the mass basis of the vector fields, i.e., based on the ``classical'' method. After the field redefinition\,(\ref{eq:mass_basis}), one obtains interactions of the dark photon in this basis as follows:
\eqs{
	\mathcal{L}_\mathrm{int} 
	& = \left(C_{VA'}\,g_V + C_{AA'}\,e Q_f \right) A'_\mu \overline f \gamma^\mu f
	+ \left(C_{VV}\,g_V + C_{AV}\,e Q_f \right) V_\mu \overline f \gamma^\mu f 
	+ C_{AA}\,e Q_f\,A_\mu \overline f \gamma^\mu f \,.
}
Here, $C_{IJ}$ denotes the element of the mixing matrix $C$ in \cref{eq:mass_basis} with $I$, $J$ labeling the vector fields $A'$, $V$ and $A$. Then, we obtain the decay rate of the dark photon into a pair of the fermions $f \bar{f}$ at tree level, assuming $f$ is a massless particle, as
\eqs{
	\Gamma(A' \to f\bar{f}) = \frac{m_{A'}}{16 \pi} \frac43 (g^{A'}_f)^2 \,, \qquad 
	g^{A'}_f \equiv C_{VA'}\,g_V + C_{AA'}\,e Q_f
}
In the following discussion, $g^{I}_f$ denotes the coupling of $f$ to the vector field $I$ in the mass basis. 

When we consider the limit of a small kinetic mixing between $A'$ and $V$ (namely, $|\eta| \ll 1$) but $\delta \to 1 - 0^+$ where $0^+$ is a small positive number, the mass mixing angle tends toward its maximal value, $\theta \simeq -\pi/4$. In this limit, the mixing matrix $C_\mathrm{kin}$ approaches the identity matrix. For the matrix elements relevant to the $A'$ decay, we obtain $C_{VA'}\simeq -1/\sqrt{2}$, and $C_{AA'}\simeq 0$ at $\mathcal{O}(\eta^0)$, while for the matrix elements relevant to the $V$ decay, we have $C_{VV}\simeq 1/\sqrt{2}$, and $C_{AV} \simeq 0$ at $\mathcal{O}(\eta^0)$. This implies that $V$ and $A'$ have the same coupling structure to fermions, and the decay rates of $A'$ and $V$ are identical in the limit of the small kinetic mixing but the degenerate mass as
\eqs{
	\Gamma(A' \to f\bar{f}) \simeq 
	\Gamma(V \to f\bar{f}) \simeq 
	\frac{m_{A'}}{16 \pi} \frac23 g_V^2 \,.
}

\subsubsection*{``Mass-insertion'' method}
\begin{figure}[t]
    \centering
    \includegraphics[width=0.6\textwidth]{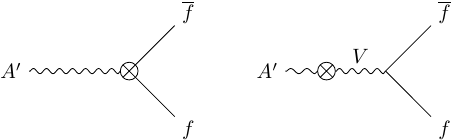}
    \caption{\small \sl
    \textbf{Left:} $A'$ decay via the kinetic mixing. \textbf{Right}: $A'$ decay via the insertion of the mass mixing.
    }
    \label{fig:MI}
\end{figure}

Next, we regard the mass mixing in \cref{eq:massmat} as a perturbation parameter and calculate the decay rate in the basis where the kinetic terms are diagonalized but the mass mixing (\ref{eq:mass_basis}) still remains.
We refer to this method for the decay rate of the dark photon as that of the ``mass-insertion'' method.
There are two diagrams that contribute to the $A'$ decay, as shown in \cref{fig:MI}: one is from the direct coupling of $A'$ to $f \bar{f}$ after diagonalizing (canonicalizing) the kinetic term (left panel), and the other is from the $A'$-$V$ mass mixing, where $V$ propagates as an intermediate state (right panel). Then, the decay rate is calculated as 
\eqs{
	&\Gamma^\mathrm{MI}(A' \to f\overline{f}) = \frac{M_{A'}}{16 \pi} \frac43 
	\left| \overline g_f^{A'} + \overline g_f^{V} \frac{- \overline m_V^2 \eta}{M_{A'}^2-\overline m_V^2+i \overline m_V \overline{\Gamma}_V} \right|^2 \,, \\
	&\overline g_f^{A'} \equiv (C_\mathrm{kin})_{VA'}\,g_V + (C_\mathrm{kin})_{AA'}\,e Q_f \,, \quad 
	\overline g_f^{V} \equiv (C_\mathrm{kin})_{VV}\,g_V + (C_\mathrm{kin})_{AV}\,e Q_f \,,
	\label{eq:rate_MI}
}
where $C_\mathrm{kin}$ denotes the diagonalizing matrix for the kinetic mixing term, $\overline g^{I}_f$ the coupling of $f$ to the vector field $I$ in the basis with the mass mixing, and $M_{A'}^2 = \overline m_V^2 (\delta^2 + \eta^2)$. $\overline \Gamma_V$ denotes the total decay rate of the particle $V$ in the absence of mixing with $A'$,
\eqs{
	\overline{\Gamma}_V = \frac{\overline m_V}{16 \pi} \frac{4}{3} g_V^2 \,.
} 

We consider the same limit as that used in the above ``classical'' method, i.e., a tiny mixing $|\eta| \ll 1$ and $\delta \to 1 - 0^+$. In this limit, the mixing matrix $C_\mathrm{kin}$ approaches the unit matrix, and its off-diagonal elements are proportional to the kinetic mixing related to $V$ and $A$. Therefore, the decay rate $\Gamma^\mathrm{MI}(A' \to \overline{f}f)$ is proportional to the square of the kinetic mixing parameter, $\eta^2$, and shows the different behavior compared to the case of the ``classical'' method.

\subsubsection*{Pole method}
\begin{figure}[t]
    \centering
    \includegraphics[width=0.95\textwidth]{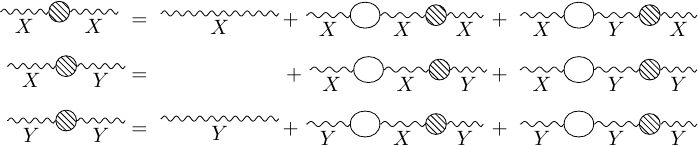}
    \caption{\small\sl
    The Dyson equations for the $X$--$X$, $X$--$Y$, and $Y$--$Y$ propagators. Here, a propagator with a shaded blob depicts the corresponding 1PI-corrected propagator, while a white blob shows the 1PI correction.
    }
    \label{fig:Pi}
\end{figure}

The result of the ``mass-insertion'' method seems inconsistent with that of the ``classical'' method at $\delta \to 1 \pm 0^+$ (though not explicitly discussed above, a similar discussion holds for $\delta \to 1 + 0^+$) . In the small mixing limit, the ``mass-insertion'' method is expected to be a good approximation for the decay rate since the mixing parameter can be regarded as a perturbative parameter. On the other hand, from the mass-insertion term in \cref{eq:rate_MI}, we see that it is no longer valid to treat the mass insertion perturbatively when $|\eta| \gtrsim \overline{\Gamma}_V/\overline m_V$ at the limit of $\delta \to 1 \pm 0^+$. In such a case, the ``classical'' method seems to provide a good approximation to calculate the decay rate. To clarify this problem, we consider the total decay rate given by the imaginary part of the propagator pole. The particles $V$ and $A'$ appear as intermediate states in scattering processes such as $f \bar{f} \to f \bar{f}$ and the location of the pole singularities of such a scattering amplitude characterizes their masses and widths. Their propagators are mixed with each other in the presence of quantum corrections since $f$ couples to both $V$ and $A'$. We denote the one-particle irreducible (1PI) vacuum polarizations involving the vector fields $X$ and $Y$ as 
\eqs{
	\Pi_{\mu\nu}^{IJ}(p^2) & = \left( g_{\mu\nu} - \frac{p_\mu p_\nu}{p^2} \right) \Pi_{IJ}(p^2) + \frac{p_\mu p_\nu}{p^2} \Delta_{IJ}(p^2) \,,
}
where $p^\mu$ is the four-momentum of the vector fields. We drop the $\Delta_{IJ}$ term in our analysis because only the first term contributes to the gauge-independent part of the corrections. The 1PI corrected propagators are obtained by solving the Dyson equations in the $I$--$J$ system shown in \cref{fig:Pi}. We denote the 1PI corrected propagator involving $I$ and $J$ as $-iD_{IJ}$, and the equations are given by
\eqs{
	-iD_{XX}(s) & = \frac{-i}{s-m_X^2 } \left[ 1 + i\Pi_{XX}(s) (-i D_{XX}(s)) + i\Pi_{XY}(s) (-iD_{YX}(s)) \right] \,, \\
	-iD_{XY}(s) & = \frac{-i}{s-m_X^2} \left[ 0 +  i\Pi_{XX}(s) (-iD_{XY}(s)) + i\Pi_{XY}(s) (-iD_{YY}(s)) \right] \,, \\
	-iD_{YY}(s) & = \frac{-i}{s-m_Y^2} \left[ 1 + i\Pi_{XY}(s) (-iD_{XY}(s)) + i\Pi_{YY}(s) (-iD_{YY}(s)) \right] \,,
}
where $s = p^2$, and $m^2_{X\,Y}$ denotes the mass eigenvalues in \cref{eq:mass_mixing}. 
Solving the equations gives
\eqs{
	D_{XX}(s) & = \frac{1}{s-m_X^2-\Pi_{XX}(s) \displaystyle - \frac{(\Pi_{XY}(s))^2}{s-m_{Y}^2 -\Pi_{YY}(s)}} \,, \\
	D_{YY}(s) & = \frac{1}{s-m_{Y}^2 -\Pi_{YY}(s) \displaystyle -\frac{(\Pi_{XY}(s))^2}{s-m_{X}^2 -\Pi_{XX}(s)}} \,, \\
	D_{XY}(s) & = \frac{\Pi_{XY}(s)}{(s-m_{Y}^2 -\Pi_{YY}(s))(s-m_V^2-\Pi_{XX}(s))-(\Pi_{XY}(s))^2} \,.
    \label{eq: propagators}
}

All of these propagators have poles at the same positions in the complex $s$ plane, which satisfies
\eqs{
	[s-m_{X}^2 -\Pi_{XX}(s)][s-m_Y^2-\Pi_{YY}(s)]-[\Pi_{XY}(s)]^2 = 0 \,.
}
Since the vacuum polarizations are generally complex-valued, pole positions are also complex-valued and thus the vector fields acquire widths. The vacuum polarizations have the form of $\Pi_{IJ}(s) = s \, \Pi_{IJ}^{\prime}(0) \,, (I \,, J = X \,, Y)$ at the one-loop level (up to the logarithmic dependence of the real part, which we ignore as discussed later), assuming $f$ is a massless fermion. Here, the prime `$\prime$' denotes a derivative with respect to the variable $s$.
In this case, we can explicitly solve the above equation for the pole positions and find two solutions as follows: 
\eqs{
	s_\mathrm{pole}^\pm & = \frac12 \frac{1}{(1-\Pi'_{YY})(1-\Pi'_{XX})-(\Pi'_{XY})^2} 
	\left\{ m_{Y}^2 (1-\Pi_{XX}') + m_{X}^2 (1-\Pi_{YY}') \right. \\
	& \qquad \left. \pm \sqrt{\left[ m_{Y}^2 (1-\Pi_{XX}') - m_{X}^2 (1-\Pi_{YY}') \right]^2 + 4 m_{Y}^2 m_{X}^2(\Pi'_{XY})^2 } \right\} \,.
	\label{eq:spole}
}
The real part of the pole rigorously defines the mass of the vector field, which is close to its tree-level value $m_{V, A'}^2$ in a weak interacting theory such as the one we discuss in this article. This is why we ignore the real part of the vacuum polarization.
The imaginary part of the pole gives the product of the mass and the total decay rate.

Here, we consider again the same limit used in the previous two methods: a tiny kinetic mixing parameter$\eta \ll 1$ and $\delta \to 1 - 0^+$. 
In this limit, as we saw in the first (``classical'') method, the couplings of $X$ and $Y$ to $f \bar{f}$ have the same magnitude but the opposite sign, and hence one obtains $\Pi_{XX}' \simeq \Pi_{YY}' \simeq - \Pi_{XY}' = \mathcal{O}(\eta^0)$. We write them as $\Pi'$ in the following discussion. The mass eigenvalues are close to each other in this limit (i.e., $m_{X}^2 \simeq m_{Y}^2$), and the mass difference is proportional to $\eta$. 
Then, when the mass difference ($\simeq \eta \, \overline m_V^2$) is negligible compared to the vacuum polarization $\Pi'$, the poles are approximately obtained at the order of $\eta^0$ as
\eqs{
	s_\mathrm{pole}^\pm \simeq m_X^2 \left( 1 + \Pi' \pm \Pi' \right) \,.
}
Thus, one of the imaginary parts of the poles is given by $2 m_X^2 \mathrm{Im} \Pi'$, i.e., at the order of $\eta^0$, while the other one is, at least, given at the order of $\eta^2$. Meanwhile, when the mass difference dominates the value inside of the square root, the poles are approximately given at the order of $\eta^0$ as,
\eqs{
	s_\mathrm{pole}^\pm \simeq m_X^2 \left( 1 + \Pi' \right) \,.
}
In this case, the imaginary parts of the poles, namely the decay rates of the two vector fields, become identical. 
The above discussion shows that the poles $s_\mathrm{pole}^\pm$ provide the behavior of the decay rate expected by the two approximation methods: One of the mass eigenstates has the total decay rate entirely suppressed at $\delta \to 1 -0^+$ when $|\eta| \lesssim |\Pi'| \simeq \overline \Gamma_V/\overline m_V$, while the decay rates of the eigenstates are identical when $|\eta| \gtrsim \overline \Gamma_V/\overline m_V$.
Note that even in this discussion the real part of $\Pi$ does not play a significant role, since it does not contribute to the mass difference in the square root and suppressed by $\pi$ compared to the imaginary part.

The particle identification of the poles might be confusing at $\delta \simeq 1$. 
The pole $s_\mathrm{pole}^+$ corresponds to $V$ at $\delta \ll 1$, while it corresponds to $A'$ at $\delta \gg 1$. 
Generally, there is a gap of order of $\eta$ between the real parts of the poles at $\delta \simeq 1$. Therefore, when we define the poles $s_\mathrm{pole}^{V}$ and $s_\mathrm{pole}^{A'}$ being those of $V$ and $A'$ as above, their real parts have a discontinuity at $\delta \simeq 1$ as a function of $\delta$. Meanwhile, the imaginary parts of the poles must be connected even at $\delta \simeq 1$: The decay rates of $V$ and $A'$ are identical at large $|\eta|$, while the decay rate of $A'$ is suppressed at small $|\eta|$. In the following, we define $s_\mathrm{pole}^{A'}$ such that its imaginary part is continuously connected to what is suppressed by $\eta^2$ apart from $\delta \simeq 1$.

\section{Dark photon decay on vector resonance} \label{sec:DP_Decay}

We discuss in this section the three methods, i.e., ``classical'', ``mass-insertion'', and ``pole'' methods to calculate the decay rate of the mediator particle in more detail using concrete examples of the degenerate partner. As expected from the discussion in \cref{sec:Med_Decay}, we will quantitatively see that when the mediator particle degenerates with an SM particle in mass, the ``classical'' and ``mass-insertion'' methods provide different decay rates and find that the total decay rate is well approximated by the ``classical'' method when $|\eta| \gtrsim \overline \Gamma_V/\overline m_V$, while by the ``mass-insertion'' method when $|\eta| \lesssim \overline \Gamma_V/\overline m_V$. This criterion is explicitly demonstrated by considering several degenerate partners, such as the $Z$ boson, the $\rho$ meson, and the true muonium. These are indeed good examples of the degenerate partner having a broad decay width (the $\rho$ meson, which has $\overline \Gamma_\rho/ \overline m_\rho \simeq 0.2$), a narrow decay width (the true muonium, which has $\overline \Gamma_{V}/ \overline m_{V} \simeq 10^{-12}$), and an intermediate decay width (the $Z$ boson, which has $\overline \Gamma_{Z}/ \overline m_{Z} \simeq 0.03$).

\subsection{Dark photon degenerate with the \texorpdfstring{$Z$}{Z} boson}

First, we consider the dark photon degenerate with the $Z$ boson. The Lagrangian is given by \cref{eq:Lagrangian_SM-DS}, and canonical kinetic terms are obtained by the following redefinitions of the gauge fields:
\eqs{
	A'_\mu & \to \frac{1}{(1 - \epsilon^2/\cos^2\theta_W)^{1/2}} A'_\mu \,, \\
	Z_\mu & \to Z_\mu - \frac{\epsilon\tan\theta_W}{(1 - \epsilon^2/\cos^2\theta_W)^{1/2}}A'_\mu \,, \\
	A_\mu & \to A_\mu + \frac{\epsilon}{(1 - \epsilon^2/\cos^2\theta_W)^{1/2}}A'_\mu \,. 
}
Here, $\theta_W$ is the Weinberg angle, again.
The mixing matrix for diagonalizing the kinetic terms of the vector fields ($A',Z,A$) is therefore given by 
\eqs{
	C_\mathrm{kin} = 
	\begin{pmatrix}
		\displaystyle \frac{1}{\left( 1 - \epsilon^2 / \cos^2{\theta_W} \right)^{1/2}} & 0 & 0 \\
		\displaystyle - \frac{\epsilon \, \tan{\theta_W}}{\left( 1 - \epsilon^2 / \cos^2{\theta_W} \right)^{1/2}} & 1 & 0 \\
		\displaystyle \frac{\epsilon }{\left( 1 - \epsilon^2 / \cos^2{\theta_W} \right)^{1/2}} & 0 & 1
	\end{pmatrix} \,.
}
Here, one obtains the mass matrix parameters $\eta$ and $\delta$, which are defined in \cref{eq:massmat}, as follows:
\eqs{
	\eta = \frac{\epsilon \, \tan{\theta_W}}{\left( 1 - \epsilon^2 / \cos^2{\theta_W} \right)^{1/2}} \,, \qquad 
	\delta = \frac{\overline{m}_{A'} / \overline{m}_Z}{\left( 1 - \epsilon^2 / \cos^2{\theta_W} \right)^{1/2}} \,.
}
The $Z$ boson decays into the SM fermions except for the top quark. 
The vector coupling of the $Z$ boson to a fermion $f$ is $g_{f} = g_Z (T_{3f} - 2 Q_f \sin^2 \theta_W)$, while the axial-vector coupling is $g_{5f} = g_Z T_{3f}$. Here, $g_Z = g/(2 \cos \theta_W)$, $T_{3f}$ is the weak isospin, and $Q_f$ is the electromagnetic charge. 
Using $\sin^2 \theta_W = 0.22$, the fine structure constant of $\alpha^{-1} = 128$, and $\overline{m}_Z = 91.19\,\mathrm{GeV}$, we obtain the SM prediction of the $Z$ boson decay width as $\overline \Gamma_Z \simeq 2.4 \, \mathrm{GeV}$, assuming $f$ is a massless fermion. The decay rate of the dark photon $A'$ using the ``classical'' method is given by 
\eqs{
	\Gamma_{A'} & = \frac{m_{A'}}{16 \pi} \frac43 \sum_f \left[ (g^{A'}_{f})^2 + (g^{A'}_{5f})^2 \right] \,, \\
	g^{A'}_{f} & = C_{ZA'}\,g_{f} + C_{AA'}\,e Q_f \,, \quad 
	g^{A'}_{5f} = C_{ZA'}\,g_{5f} \,.
}
Here, $m_{A'}$ is the mass eigenvalue given in \cref{eq:mass_mixing}.

As mentioned in the previous section, the above decay rate approaches $\overline \Gamma_Z/2$ at $\delta \to 1\pm0^+$, as the mixing matrix element approaches $C_{ZA'} \simeq - 1/\sqrt{2}$. Meanwhile, the coupling of the dark photon $A'$ to the SM fermion through mixing with the $Z$ boson is proportional to the matrix element $C_{ZA'}$, which is approximated at $\eta \ll |1 - \delta^2|$ as 
\eqs{
	C_{ZA'} = (C_\mathrm{kin})_{ZA'} \cos\theta - \sin \theta  
	\simeq - \eta - \left( \frac{- \eta}{1-\delta^2} \right) \,.
}
It is found that the $A'$ coupling through mixing with the $Z$ boson, $C_{ZA'}$, is suppressed at $\delta \simeq 0$. As seen in \cref{sec:truemuonium,app:rhoee}, it happens at a different $\delta$ in other examples, e.g., $\delta \sim 1$.

The decay rate of the dark photon $A'$ using the ``mass-insertion'' method is 
\eqs{
	\Gamma^\mathrm{MI}_{A'} & = 
	\frac{M_{A'}}{16 \pi} \frac43 \sum_ f (|\overline g_{f}|^2 + |\overline g_{5f}|^2) \,, \\
    \overline g_{f} & = \overline g_{f}^{A'} + \overline g_{f}^Z \frac{- \eta \overline m_Z^2}{M_{A'}^2-\overline m_Z^2+i \overline m_Z \overline \Gamma_Z} \,, \quad 
	\overline g_{5f} = \overline g_{5f}^{A'} + \overline g_{5f}^Z \frac{- \eta \overline m_Z^2}{M_{A'}^2-\overline m_Z^2+i \overline m_Z \overline \Gamma_Z} \,.
	\label{eq:Zrate_MI}
}
Here, $M_{A'}^2 = \overline{m}_Z^2 (\delta^2 + \eta^2)$, and $\overline \Gamma_Z$ is the decay rate of the SM $Z$ boson in the absence of $A'$,
\eqs{
	\overline \Gamma_Z = \frac{\overline{m}_{V}}{16 \pi} \frac43 \sum_f \left( g_{f}^2 + g_{5f}^2 \right) \,.
}
The vector and axial-vector couplings to the $Z$ boson and the dark photon in the basis with the mass mixing are 
\begin{align}
    \overline g_{f}^{A'} & \equiv (C_\mathrm{kin})_{ZA'}\,g_{f} + (C_\mathrm{kin})_{AA'}\,e Q_f \,, &
	\overline g_{f}^{Z} & \equiv (C_\mathrm{kin})_{ZZ}\,g_{f} + (C_\mathrm{kin})_{AZ}\,e Q_f \,, \nonumber \\
    \overline g_{5f}^{A'} & \equiv (C_\mathrm{kin})_{ZA'}\,g_{5f} \,, &
	\overline g_{5f}^{Z} & \equiv (C_\mathrm{kin})_{ZZ}\,g_{5f} \,.
\end{align}
Since the mass-insertion contribution cancels with the contribution from the direct coupling as $M_{A'}^2 \to 0$, contributions from the $Z$-boson couplings $g_{5 f}$ are suppressed. 
We find similar behavior in the ``classical'' method, but the cancellation gets gentle because of the presence of the imaginary part $i \overline{m}_Z \overline \Gamma_Z$. 
The ``mass-insertion'' method is expected to be no longer valid as $|\eta| \simeq |\epsilon| \gtrsim \overline \Gamma_Z/\overline{m}_Z \simeq 0.03$.

\begin{figure}[t]
    \centering
    \includegraphics[width=0.95\textwidth]{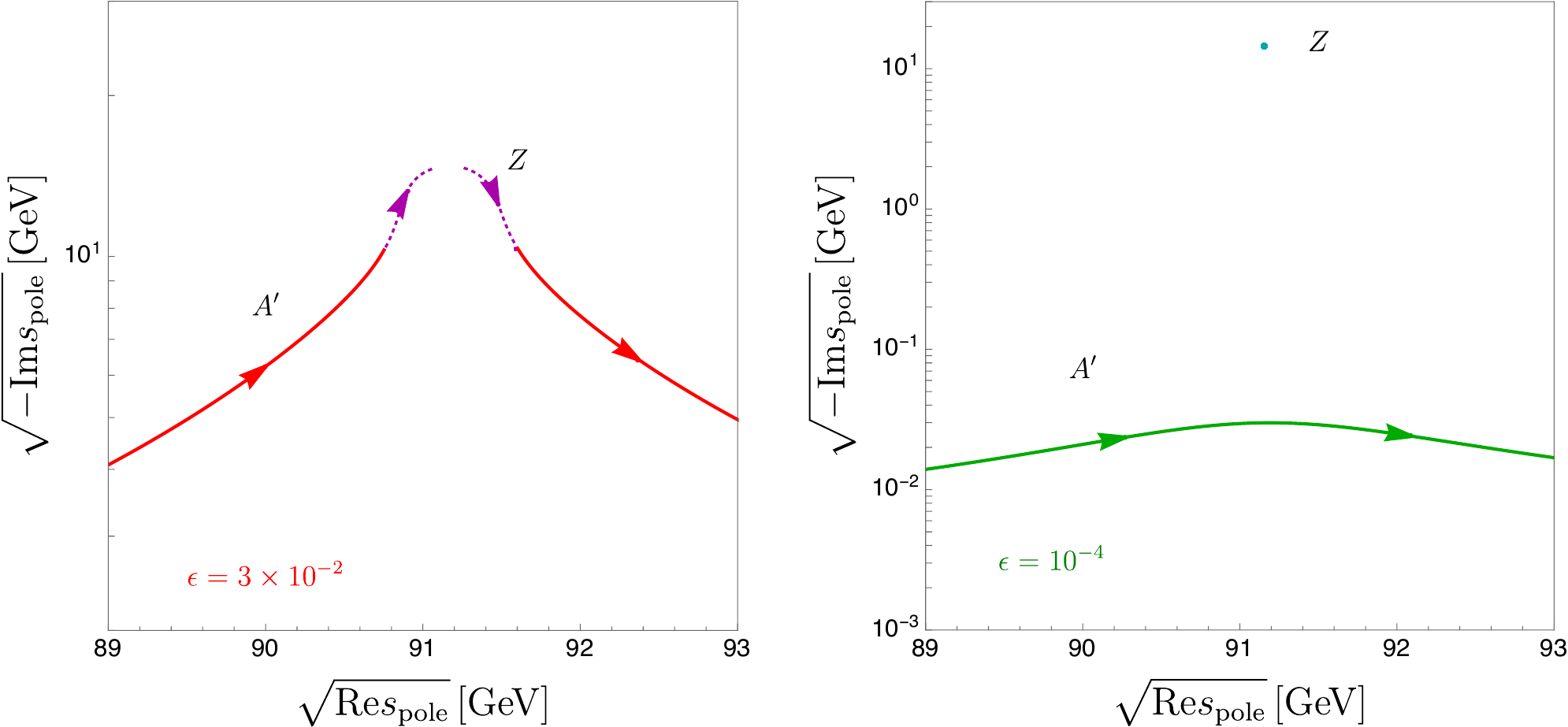}
    \caption{\small\sl
    Parametric plots of the real and imaginary parts of the propagator poles as a function of the parameter $\delta \in [0.95, 1.05]$, with $\epsilon$ being fixed to be  $3 \times 10^{-2}$ (left panel) and $10^{-4}$ (right panel), respectively.
    }
    \label{fig:Zpole}
\end{figure}

The vacuum polarization rigorously determines the total decay rate in terms of the propagator pole. The imaginary part of the vacuum polarization involving the vector fields $I$ and $J$ is 
\eqs{
	\mathrm{Im} \Pi_{IJ}(s) 
	& = - \frac{\pi}{16 \pi^2} \frac{4}{3} s \sum_{f} (g_{f}^{I} g_{f}^{J} + g_{5f}^{I} g_{5f}^{J}) \,,
}
where $g_{f}^{I}$ and $g_{5f}^{I}$ are vector and axial-vector couplings of the fermion $f$ to the vector field $I$ in the mass basis. \cref{fig:Zpole} shows the parametric plots of the real and imaginary parts of the pole as a function of the parameter $\delta$. The real part of the $Z$ pole changes from $\overline m_Z^2 (1 + \eta)$ to $\overline m_Z^2 (1 - \eta)$ along the change of $\delta  \in [0.95, 1.05]$, while the real part of the $A'$ pole is proportional to $\delta^2$. The imaginary parts of both the poles are identical near $\delta = 1$ when $\epsilon = 3 \times 10^{-2}$ (left panel). Meanwhile, when $\epsilon = 10^{-4}$ (right panel), the $Z$ pole is almost unchanged, while the imaginary part of the $A'$ pole enhances at $\delta \simeq 1$.
Note that the real part of the pole will receive a correction from the real part of the vacuum polarization, which we ignore.
We focus on the imaginary part in the followings.

\begin{figure}[t]
    \centering
    \includegraphics[width=0.46\textwidth]{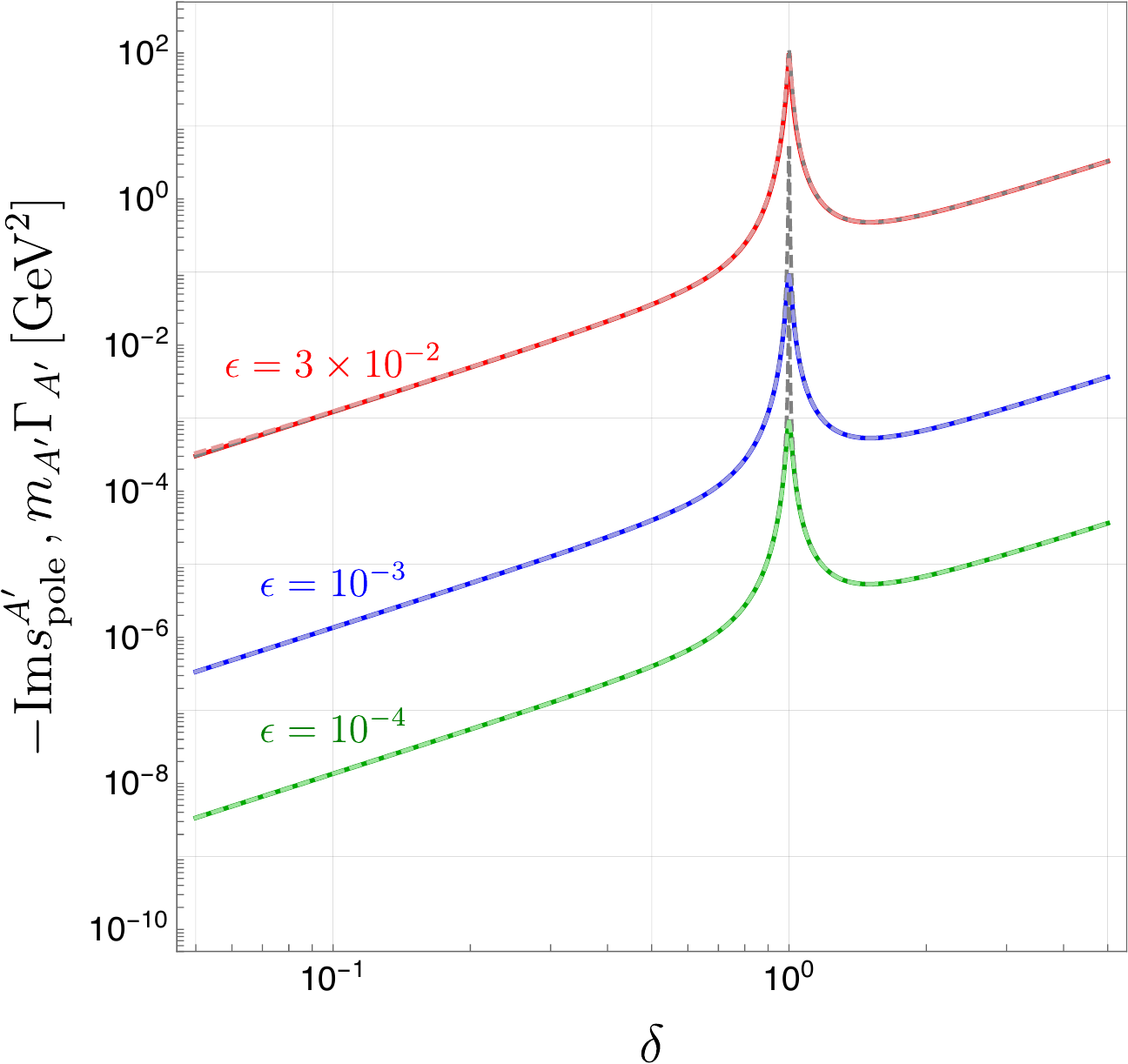}
    \qquad
    \includegraphics[width=0.46\textwidth]{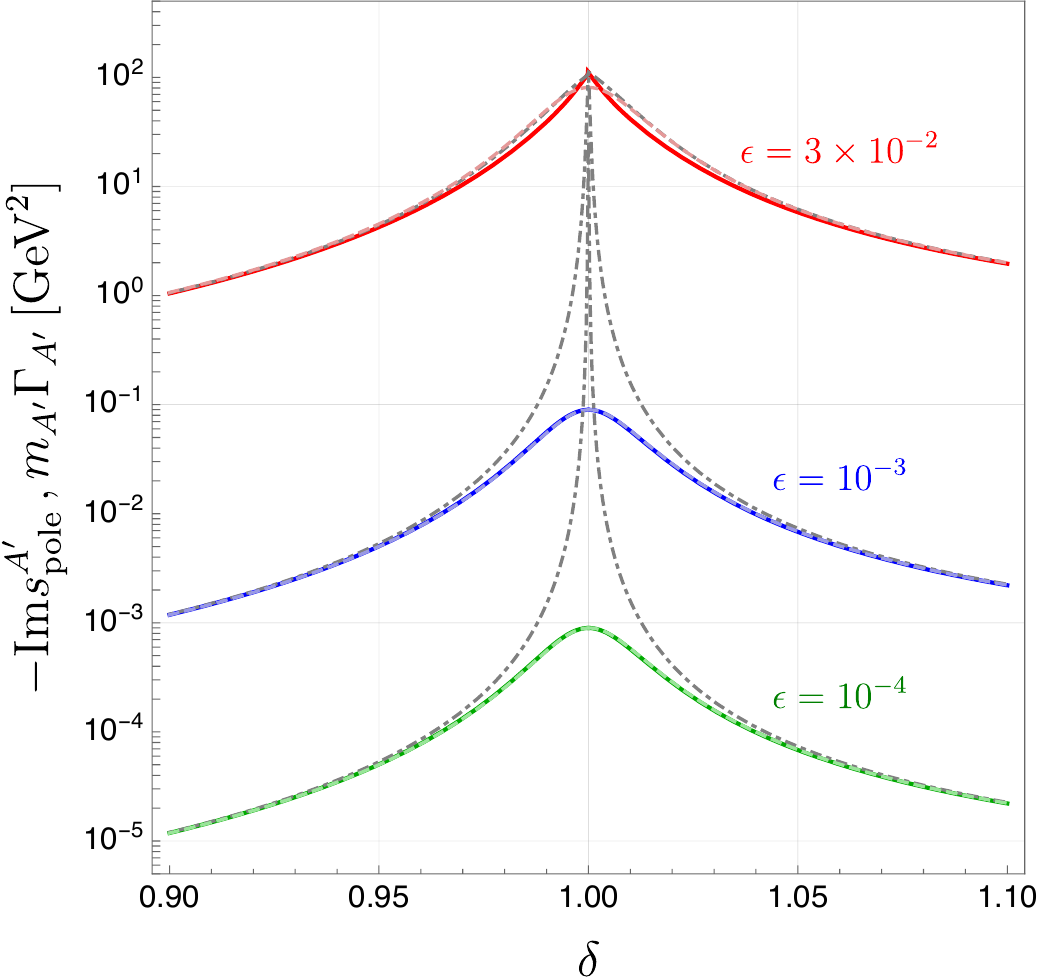}
    \caption{\small\sl
    \textbf{Left:} The $A'$ decay rates calculated the three methods for various values of $\epsilon$: the imaginary part of the propagator pole (colored solid lines), the ``mass-insertion'' method (colored dashed lines), and the ``classical'' method (gray dot-dashed lines). \textbf{Right:} A closer look at the decay rates near $\delta = 1$.
    }
    \label{fig:Zrate}
\end{figure}

The decay rate computed by the three methods is depicted in \cref{fig:Zrate} as a function of $\delta$, with several choices of $\epsilon$. 
The right panel is a closer look at the decay rate at $\delta \sim 1$, and the color code and the line type are the same as the left one.  
First, it is seen from the figure that the three methods give the same rate except for the region of $\delta \sim 1$. Next, when the mass difference $|1-\delta|$ is smaller than $\overline \Gamma_Z/\overline{m}_Z \simeq 3 \times 10^{-2}$, the result of the ``classical'' method (denoted by dot-dashed lines) is not compatible with that of the ``mass-insertion'' method (denoted by dashed lines). Third, the result of the ``mass-insertion'' method agrees with the decay rate obtained by the imaginary part of the pole method (denoted by solid lines) when $\epsilon$ is sufficiently small. On the other hand, the result of the ``classical'' method does not agree with the rate obtained by the pole method because it approaches $\overline \Gamma_Z/2$ at $\delta \simeq 1$ in the ``classical'' method. When $\epsilon$ becomes larger than $\overline \Gamma_Z/\overline{m}_Z$, as expected from the discussion in the previous section, the result of the ``mass-insertion'' method is no longer in agreement with that obtained by the pole, which is seen in the lines for $\epsilon = 3 \times 10^{-2}$.

\begin{figure}[t]
    \centering
    \includegraphics[width=0.5\textwidth]{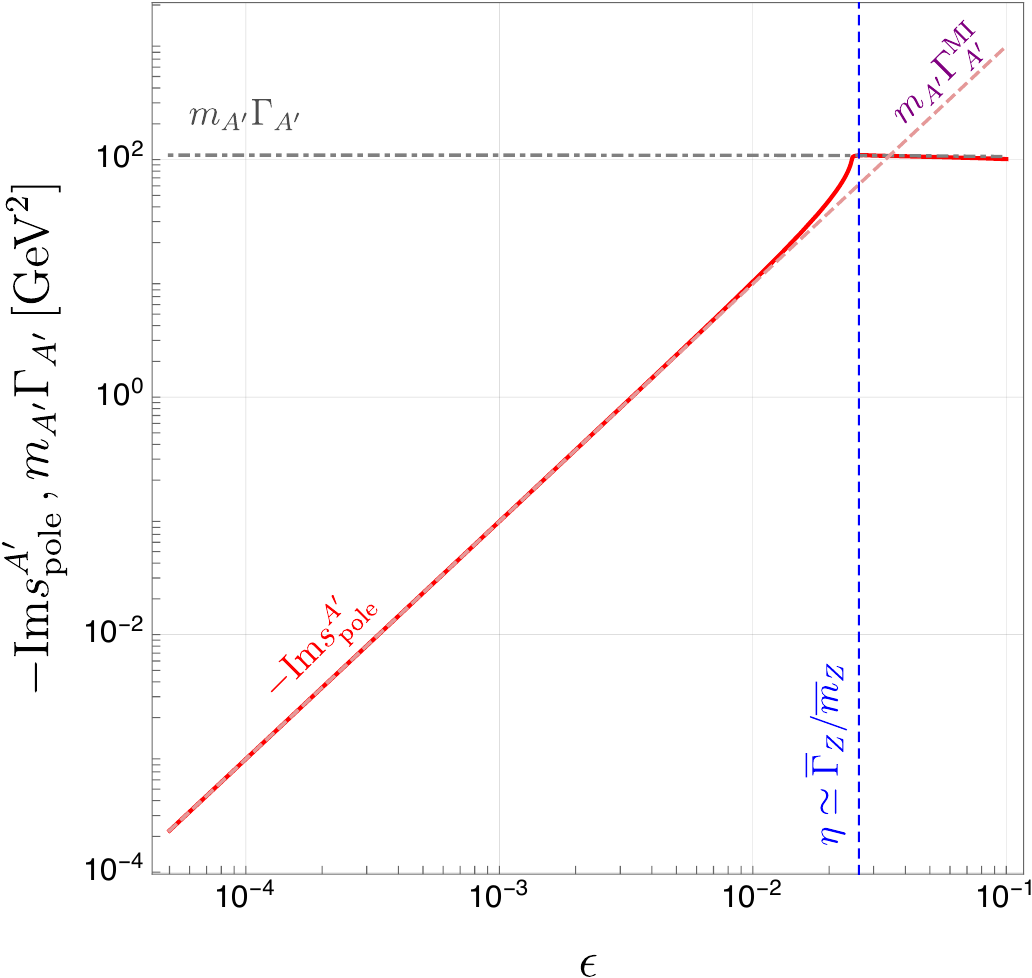}
    \caption{\small\sl
    The decay rates of the dark photon $A'$ degenerate with $Z$ in mass calculated in the three methods as a function of $\epsilon$ with $\delta = 1$: the imaginary part of the propagator pole (red solid line), the ``mass-insertion'' method (pink dashed line), and the ``classical'' method (gray dot-dashed line). The blue-dashed vertical line indicates the boundary where the ``mass-insertion'' method is no longer valid. 
    }
    \label{fig:Zrate_eps}
\end{figure}

We also show in \cref{fig:Zrate_eps} the comparison of the decay widths obtained by the three methods as a function of $\epsilon$, with $\delta$ being fixed to be one. It is seen from the figure that the decay width obtained by the ``mass-insertion'' method accurately reproduces the width obtained by the imaginary part of the propagator pole when $|\epsilon| \lesssim \overline \Gamma_Z/ \overline m_Z$. On the other hand, the decay width obtained by the ``classical'' method accurately reproduces the width obtained by the propagator pole when $|\epsilon| \gtrsim \overline \Gamma_Z/ \overline m_Z$.

\subsection{Dark photon degenerate with the \texorpdfstring{$\rho$}{rho} meson}

We consider the case where the dark photon is degenerate with the $\rho$ meson, one of the vector hadronic resonances.%
\footnote{
    We also discuss the SM example, a $\rho$--$\omega$--$\gamma$ system, in \cref{app:rhoomegagamma}.
} 
We introduce the vector meson as a gauge boson of the hidden local symmetry\,\cite{Bando:1984ej}. (See Ref.\,\cite{Harada:2003jx} for a review.) 
The low-energy Lagrangian of the two-flavor QCD is given as follows: 
\eqs{
	\mathcal{L} = &
	- \frac14 F_{\mu\nu} F^{\mu\nu} 
	- \frac14 F'_{\mu\nu} F'^{\mu\nu} 
	- \frac14 V^a_{\mu\nu} V^{a \mu\nu}
	+ \frac{\epsilon}{2} F'_{\mu\nu} F^{\mu\nu} 
	- \frac{e}{g} \mathrm{tr}(Q V_{\mu\nu}) F^{\mu\nu} \\
	& + \frac12 \widetilde m_{A'}^2 A'_\mu A'^\mu 
	+ \frac12 \widetilde m_{\rho}^2 V^a_\mu V^{a\mu} 
	+ \mathcal{L}_\mathrm{int}
	+ \cdots \,.
	\label{eq:LE_Had-DPLag}
}
Here, $e$ is the electromagnetic coupling, and $g$ is the hadronic coupling. Interaction terms among the $\rho$ meson field (i.e., $V_\mu = V^a_\mu t^a$ with $t^a$ being the $SU(2)$ generator) and pseudo-Nambu-Goldstone boson (pions) are included in $\mathcal{L}_\mathrm{int}$. The charge matrix $Q$ and the vector meson matrix $V_\mu$ are given as follows:
\eqs{
	Q = \frac13 
	\begin{pmatrix}
		2 & \\ 
		& -1
	\end{pmatrix} \,, 
	\qquad 
	V_\mu = \frac{1}{2} 
	\begin{pmatrix}
		\rho_\mu^0 & \sqrt{2} \rho_\mu^+ ~~\\ 
		\sqrt{2} \rho_\mu^- ~~ & - \rho_\mu^0
	\end{pmatrix} \,.
}
As the Lagrangian shows, the SM photon kinematically mixes with both the dark photon and the neutral $\rho$ meson. There are several ways to redefine the vector fields to make their kinetic terms normalized canonically.%
\footnote{
    The kinetic mixing between the $\rho$ meson and the photon is removed by shifting the $\rho$ meson field in a conventional vector meson dominance. 
    The photon couples to the hadronic current only through mixing with $\rho$ for such a case.
} 
Here, to align our discussion with that of \cref{sec:Med_Decay}, we first remove all the kinetic mixing terms by the shift: $A_\mu \to A_\mu + \epsilon A'_\mu - (e/g) \rho^0_\mu$. This field redefinition allows us to use the same parametrization discussed in the previous section. After this shift, there is the kinetic mixing between the neutral $\rho$ meson and the dark photon and we arrive at the interaction basis. By taking the canonical normalization for the vector fields on this basis, one obtains
\eqs{
	\mathcal{L} = 
	- \frac14 F_{\mu\nu} F^{\mu\nu} 
	- \frac14 F'_{\mu\nu} F'^{\mu\nu} 
	- \frac14 \rho^0_{\mu\nu} \rho^{0 \mu\nu}
	- \frac{\epsilon_\mathrm{eff}}{2}\rho^0_{\mu\nu} F'^{\mu\nu} 
        + \frac12 \overline m_{A'}^2 A'_\mu A'^\mu 
	+ \frac12 \overline m_{\rho}^2 \rho^0_\mu \rho^{0\mu} 
	+ \mathcal{L}_\mathrm{int}
	+ \cdots \,.
	\label{eq:L_rho_Ap_mixing}
}
Here, $\overline m_{A'}^2 = \widetilde m_{A'}^2/(1-\epsilon^2)$ and $\overline m_{\rho}^2 = \widetilde m_{\rho}^2/(1-e^2/g^2)$.
The kinetic mixing term between the neutral $\rho$ meson and the dark photon is given by the product of two kinetic mixing factors, one between $A'$ and $A$ and the other between $\rho$ and $A$, as follows:  
\eqs{
	\epsilon_\mathrm{eff} = \frac{\epsilon}{(1-\epsilon^2)^{1/2}} \frac{e/g}{(1-e^2/g^2)^{1/2}} \,.
}
This kinetic mixing term is removed by the shift $\rho^0_\mu \to \rho^0_\mu + \epsilon_\mathrm{eff} A'_\mu $ and the subsequent rescaling of $A'$. Then, the parameters $\eta$ and $\delta$ are given
\eqs{
	\eta = \frac{\epsilon_\mathrm{eff}}{(1 - \epsilon_\mathrm{eff}^2)^{1/2}} \,, \qquad
	\delta = \frac{\overline m_{A'}/\overline m_\rho}{(1 - \epsilon_\mathrm{eff}^2)^{1/2}} \,,
    \label{eq: eta and delta in rho}
}
and one obtains the mixing matrix, which diagonalizes all the kinetic terms of the vector fields $(A', \rho, A)$ as
\eqs{
	C_\mathrm{kin} = 
	\begin{pmatrix}
		\displaystyle \frac{1}{(1 - \epsilon_\mathrm{eff}^2)^{1/2}} \frac{1}{(1-\epsilon^2)^{1/2}} & 0 & 0 \\
		\displaystyle - \frac{\epsilon_\mathrm{eff}}{(1 - \epsilon_\mathrm{eff}^2)^{1/2}} \frac{1}{(1-e^2/g^2)^{1/2}} & \displaystyle \frac{1}{(1 - e^2/g^2)^{1/2}} & 0 \\
		\displaystyle \frac{1}{(1 - \epsilon_\mathrm{eff}^2)^{1/2}} \frac{\epsilon}{(1-\epsilon^2)^{1/2}} + \frac{\epsilon_\mathrm{eff}}{(1 - \epsilon_\mathrm{eff}^2)^{1/2}} \frac{e/g}{(1 - e^2/g^2)^{1/2}} & \displaystyle - \frac{e/g}{(1 - e^2/g^2)^{1/2}} & 1
	\end{pmatrix} \,.
	\label{eq:Cmix_rho}
}
The $\rho$ meson predominantly decays into $\pi^+ \pi^-$, so we focus only on this channel.%
\footnote{
    We also consider a hypothetical situation in \cref{app:rhoee} that $\rho$ decays only into $e^-e^+$. In such a case, the decay width is suppressed at ${\cal O}(1)$\,keV, and it can be used to verify our discussion in a different property of the vector fields.
} Then, interactions among the vector fields and the pion fields in the interaction basis are given by
\eqs{
	\mathcal{L}_\mathrm{int} 
	\supset & - i (e A_\mu + g \rho^0_\mu) \left[ \pi^+ (\partial^\mu \pi^-) - (\partial^\mu \pi^+) \pi^- \right] + (e A_\mu + g \rho^0_\mu)^2 \pi^+ \pi^- \,.
}
One obtains interaction terms in the mass basis using the mixing matrix $C$ in \cref{eq:mass_basis}, where it is given by the product of \cref{eq:Cmix_rho} and the one diagonalizing the mass matrix described by the parameters in \cref{eq: eta and delta in rho}. The decay rate of the $\rho$ meson in the absence of $A'$ is calculated at the tree-level as
\eqs{
	\overline \Gamma_\rho
	= \frac{\overline m_\rho}{16 \pi} \frac13 g^2 \,.
    \label{eq:barGammarho}
}
Here, we assume that pions are massless in the final state.
Choosing phenomenological favorable values $g = 5.92$ and $\overline m_\rho = 775.26$\,MeV, one obtains the numerical value of $\overline \Gamma_{\rho}$ being about 180\,MeV.

Now, we calculate the decay widths of the dark photon using the three methods introduced in \cref{sec:Med_Decay}. The decay rate in the ``classical'' method, i.e., the decay rate in the mass basis, is 
\eqs{
	\Gamma_{A'}
	= \frac{m_{A'}}{16 \pi} \frac13 (g_\pi^{A'})^2 \,, \qquad 
	g_\pi^{A'} = C_{\rho A'}\,g + C_{AA'}\,e \,.
}
Note that $g_\pi^{A'}$ is approximately proportional to $\sin \theta$ with $\theta$ being the mixing angle between $A'$ and $\rho$ in contrast to the $Z$ boson case where both $\sin \theta$ and $\cos \theta$ terms survive.
Meanwhile, the decay rate in the ``mass-insertion'' method is obtained in the basis where the mass mixing between the $\rho$ meson and the dark photon $A'$ remains after removing the kinetic mixing term in \cref{eq:L_rho_Ap_mixing}. There are two contributions to the $A'$ decay, from the direct coupling of $A'$ and from the mass mixing between $\rho$ and $A'$. 
The decay width in this method is obtained as
\eqs{
	\Gamma^\mathrm{MI}_{A'}
	& = \frac{M_{A'}}{16 \pi} \frac13 \left| 
	\overline g_{\pi}^{A'} + \overline g_{\pi}^{\rho} \frac{- \eta \overline m_\rho^2}{M_{A'}^2-\overline m_\rho^2+i \overline m_\rho \overline \Gamma_\rho} \right|^2 \,, \\
	\overline g_{\pi}^{A'} & \equiv (C_\mathrm{kin})_{\rho A'}\,g + (C_\mathrm{kin})_{AA'}\,e \,, \quad 
	\overline g_{\pi}^{\rho} \equiv (C_\mathrm{kin})_{\rho \rho}\,g + (C_\mathrm{kin})_{A\rho}\,e \,.
}
Here, $M_{A'}^2 = \overline m_\rho^2 (\delta^2 + \eta^2)$. Since the interaction terms of the vector bosons have a structure similar to the standard vector meson dominance, it leads to $\overline g_{\pi}^{A'} = 0$. 
In contrast to the $Z$ boson case, there is no cancellation between the direct and the mass-insertion contributions. 
Hence, the decay rate is suppressed in two limits; it is proportional to $M_{A'}$ at small $M_{A'}$, while it is proportional to $M_{A'}^{-3}$ at large $M_{A'}$. 
The ``mass-insertion'' method is expected to be valid at $|\eta| \lesssim \overline \Gamma_\rho/\overline m_\rho \simeq 0.2$.

\begin{figure}[t]
    \centering
    \includegraphics[width=0.46\textwidth]{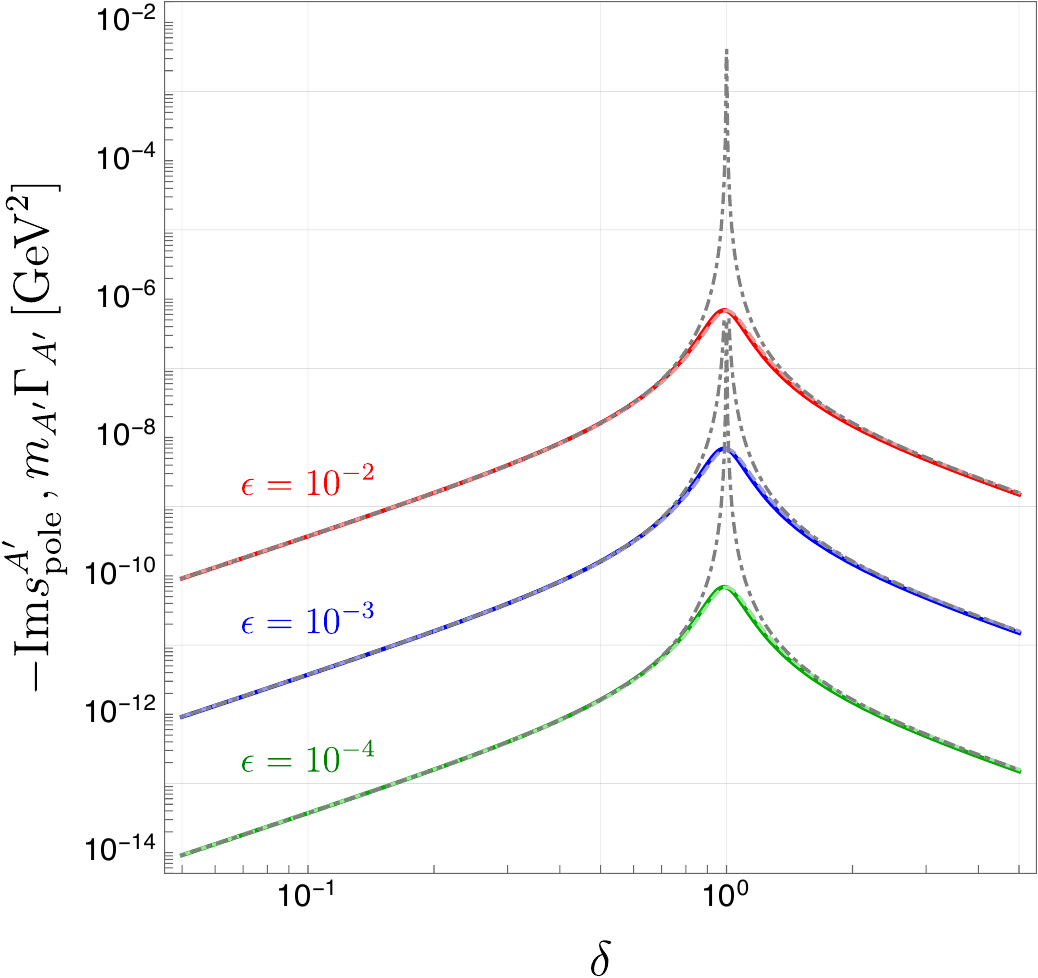}
    \qquad
    \includegraphics[width=0.46\textwidth]{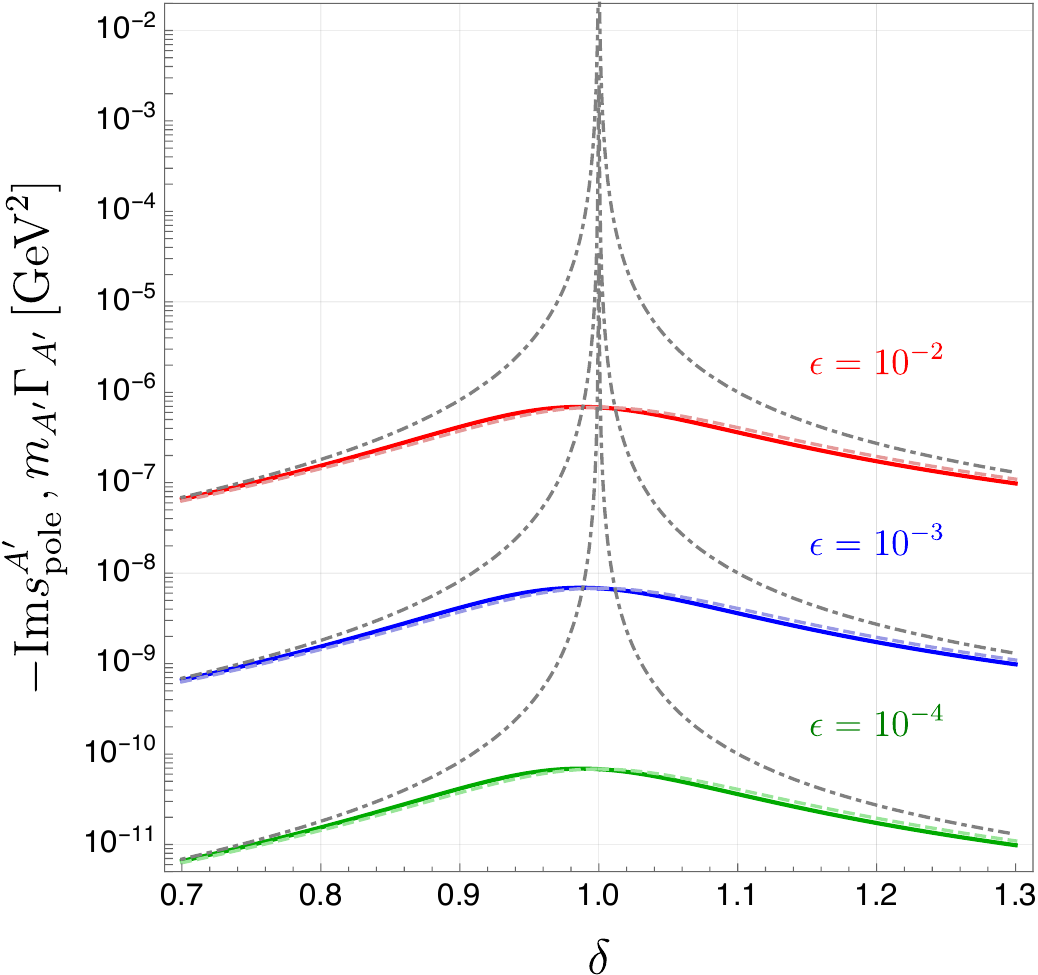}
    \caption{\small\sl
		The same as \cref{fig:Zrate}, but for $A'$ degenerate with $\rho$ in mass. Each line code is the same as in \cref{fig:Zrate}. \textbf{Left:} The decay rates as a function of $\delta$ with fixed $\epsilon$.
		\textbf{Right:} A closer look at the rates at $\delta \simeq 1$.
    }
    \label{fig:rhorate}
\end{figure}

Next, we calculate the decay width of the dark photon using the vacuum polarization for the vector fields $X$ and $Y$ at the one-loop level. There are two contributions to the vacuum polarization: one is from the three-point vertices $X\pi\pi$ and $Y\pi\pi$, and the other is from the four-point vertices, such as $XY\pi\pi$. The imaginary part of the vacuum polarization is obtained as
\eqs{
	\mathrm{Im} \Pi_{IJ} (s)
	& =
	- \frac{\pi s}{16 \pi^2} \frac{1}{3} g^I_{\pi} g^J_{\pi} \,,
}
where $g^{I}_{\pi}$ denotes the gauge coupling constant of the pions with regard to $I$ in the mass basis.

\begin{figure}[t]
    \centering
    \includegraphics[width=0.5\textwidth]{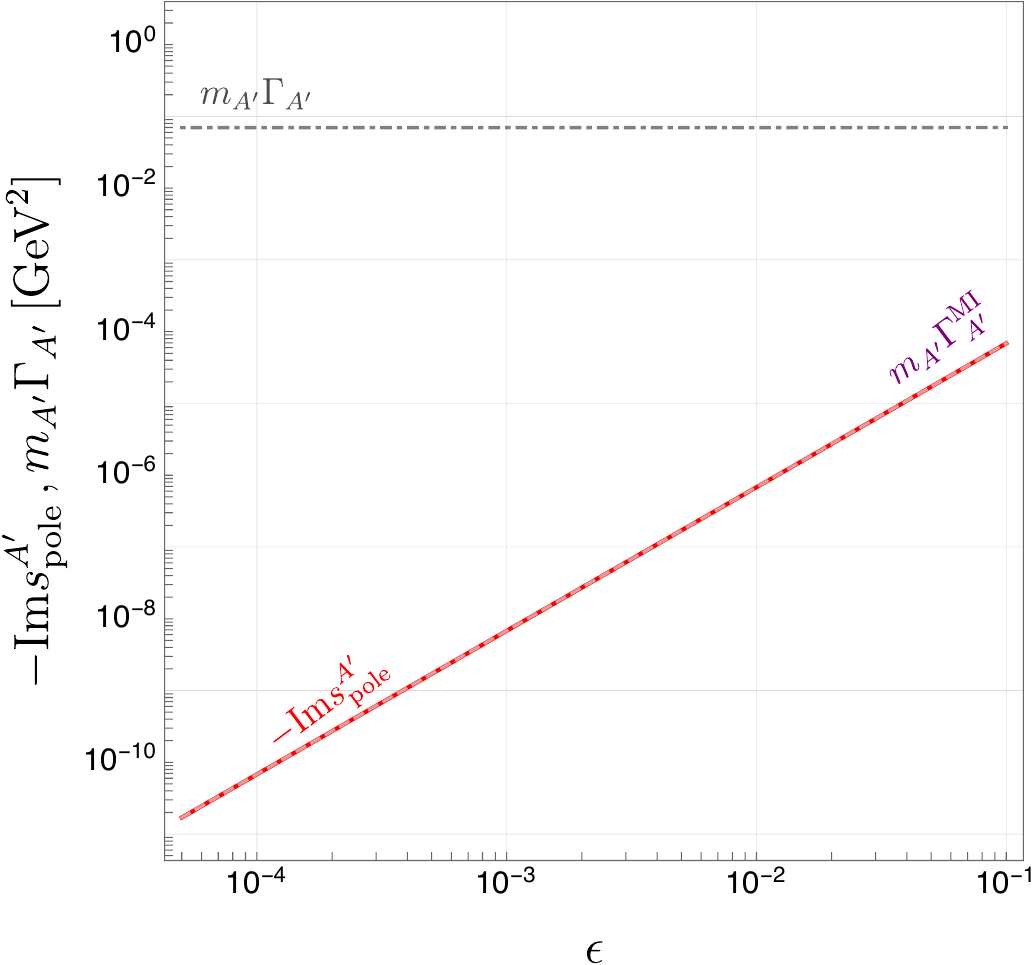}
    \caption{\small\sl
		The same as \cref{fig:Zrate_eps}, but for $A'$ degenerate with $\rho$ in mass. Each line code is the same as before.
    }
    \label{fig:rhorate_eps}
\end{figure}

\cref{fig:rhorate,fig:rhorate_eps} compares the decay rates obtained by the three methods, as those in \cref{fig:Zrate,fig:Zrate_eps}. We plot the decay rates as a function of $\delta$ in \cref{fig:rhorate} for given $\epsilon = 10^{-2}$, $ 10^{-3}$, and $ 10^{-4}$, while the decay rates as a function of $\epsilon$ in \cref{fig:rhorate_eps} with $\delta$ being fixed to be one. We use the same line types as those in \cref{fig:Zrate,fig:Zrate_eps}. 
First, as in the previous $Z$ boson case, the decay widths obtained by the three methods agree with each other except for the $\delta \sim 1$ region, and the decay width obtained by the ``classical'' method deviates from the others in the region of $|\delta - 1| \lesssim \overline \Gamma_\rho/ \overline{m}_\rho$. Next, contrary to the $Z$ boson case, the decay rates are suppressed at large $\delta$. The dark photon $A'$ does not directly couple to the pion current after diagonalizing the kinetic mixing between the $\rho$ meson and $A'$. In other words, $A'$ couples to the pion current only via the mass mixing with $\rho$. Therefore, the $A'$ coupling in the mass basis is suppressed by the mixing angle $\theta$ defined in \cref{eq:mass_mixing} at large $\delta$. 
As shown in the right panel of \cref{fig:rhorate}, the decay width with the ``mass-insertion'' method does very slightly deviate from the decay width in the pole method. This is because the strong coupling $g$ leads to sizable corrections to the real part of poles even from the imaginary parts of the vacuum polarization.
Finally, as seen in \cref{fig:rhorate_eps}, the ``classical'' method is not consistent with the other methods at $\delta \simeq 1$ even for $\epsilon \sim 10^{-1}$. The kinetic mixing between $A'$ and $\rho$ is given by the product of $\epsilon$ and $e/g$, hence the criterion $|\eta| \lesssim \overline \Gamma_\rho/ \overline m_\rho$ for the validity of the use of the ``mass-insertion'' method implies that $|\epsilon| \lesssim \overline \Gamma_\rho/ \overline m_\rho \times (e/g)^{-1} \simeq 4$.
As far as the kinetic mixing parameter is perturbative, the result of the ``mass-insertion'' method always approximates the decay width obtained by the imaginary part of the propagator pole, while the ``classical'' method always overestimates the decay width.

\subsection{Dark photon degenerate with the true muonium \label{sec:truemuonium}}

In this subsection, as the last example to demonstrate how well our criterion works, we discuss the case with the dark photon $A'$ degenerate with the true muonium in mass. 
The true muonium is the bound state composed of a muon and an antimuon.
Since the true muonium constructed with the total spin of $S = 1$ and the orbital angular momentum of $L = 0$ has the quantum numbers of $J^\mathrm{PC} = 1^{--}$, the dark photon mixes with the true muonium when its mass is around the muon threshold, i.e., the twice the muon mass.%
\footnote{
    The ortho-positronium is also a bound state with the same quantum numbers as $A'$; it decays into three photons at the one-loop level. 
    We consider the true muonium to simplify the discussion, which decays into $e^-e^+$ at the tree level.
}  
We will focus on the mixing between the dark photon and the ground state of the true muonium, which we refer to as ``true muonium'' in the following discussion. 
The low energy effective Lagrangian that describes interactions among the SM photon $A_\mu$, the ``true muonium'' $V_\mu$, and the dark photon $A'_\mu$ is given as follows:
\begin{align}
    \mathcal{L} =
    &-\frac{1}{4}F_{\mu \nu}F^{\mu \nu}
    -\frac{1}{4}F'_{\mu \nu}F'^{\mu\nu}
    -\frac{1}{4}V_{\mu\nu}V^{\mu \nu}
    +\frac{\epsilon}{2}F_{\mu \nu} F'^{\mu \nu}
    -\frac{\kappa}{2}F_{\mu \nu}V^{\mu \nu}
    \nonumber \\ 
    &+\frac{1}{2}\widetilde{m}_{A'}^2 A'_\mu A'^{\mu}
    +\frac{1}{2}\widetilde{m}_V^2V_\mu V^\mu
    +e A_\mu J^\mu_\mathrm{EM}
    +\cdots \,,
    \label{eq:lagrangian true muonium}
\end{align}
where $\widetilde m_V = 2m_\mu - \alpha^2m_\mu/4 $ is the mass parameter of the ``true muonium''. 
The effective Lagrangian \cref{eq:lagrangian true muonium} is derived from the original Lagrangian \cref{eq:Lagrangian_SM-DS} using the so-called potential non-relativistic\,(NR) Lagrangian method: it is obtained from the original Lagrangian by the NR expansion of the muon field, integrating out a soft photon field, and introducing an auxiliary field describing the ``true muonium''\,\cite{Matsumoto:2022ojl}. 
We introduce the kinetic mixing term between the SM photon and the ``true muonium'' to make the gauge invariance of the Lagrangian manifest instead of the mass mixing term between them. 
The NR method gives the mixing parameter $\kappa = \alpha^2/2$, which is also obtained through the consistency of the ``true muonium'' decay width in the absence of $A'$, $\overline \Gamma_V \simeq \alpha^5 m_\mu/6$.%
\footnote{
    We include only the contribution from the $\mu^- \mu^+$ pair annihilation into $e^- e^+$ inside the true muonium and ignore the contribution from the decay of the component particles $\mu^-$ and $\mu^+$. 
    This is because the latter contribution, $\Gamma_\mu \simeq 2.99 \times 10^{-16}$\,MeV, is much smaller than the former one $\overline \Gamma_V \simeq 3.66 \times 10^{-10}$\,MeV.
}
See \cref{app:muonium_detail} for the determination of the mixing parameter $\kappa$.

Now, we calculate the decay rates of the dark photon using the three methods discussed in \cref{sec:Med_Decay}. 
Similarly to the previous examples, i.e., the $Z$ boson and the $\rho$ meson cases, we first calculate the mixing matrix $C$ among the vector fields defined in \cref{eq:mass_basis}: removing all kinetic mixing terms from the Lagrangian (\ref{eq:lagrangian true muonium}) by shifting the SM photon field as $A_\mu \to A_\mu + \epsilon A'_\mu - \alpha^2 V_\mu/2$, and rescaling the dark photon and the ``true muonium'' as $A'_\mu \to A'_\mu/(1-\epsilon^2)^{1/2}$ and $V_\mu \to V_\mu/(1-\alpha^4/4)^{1/2}$ to make their kinetic terms canonical. Then, we arrive at the Lagrangian,
\begin{align}
    \mathcal{L} =&
    -\frac{1}{4}F_{\mu \nu} F^{\mu \nu}
    -\frac{1}{4}F'_{\mu \nu}F'^{\mu \nu}
    -\frac{1}{4}V_{\mu \nu} V^{\mu \nu}
    -\frac{\epsilon_\mathrm{eff}}{2}F'_{\mu \nu} V^{\mu \nu} \nonumber \\ 
    &
    +\frac{1}{2}\overline{m}^2_{A'}A'_\mu A'^\mu
    +\frac{1}{2}\overline{m}^2_{V}V_\mu V^\mu
    +e A_\mu J^\mu_\mathrm{EM}
    +\epsilon e A'_{\mu} J^{\mu}_\mathrm{EM}
    -\frac{e\alpha^2}{2} V_\mu J^\mu_\mathrm{EM}
    +\cdots \,.
\end{align}
Here, we introduced rescaled mass parameters $\overline{m}_{A'} = \widetilde{m}_{A'}/(1-\epsilon^2)^{1/2}$ and $\overline{m}_V = \widetilde{m}_V/(1-\alpha^4/4)^{1/2}$.
There appears a new kinetic mixing term between the dark photon and the ``true muonium'' with the mixing parameter, 
\begin{align}
    \epsilon_\mathrm{eff} =
    \frac{\epsilon}{(1-\epsilon^2)^{1/2}}
    \frac{\alpha^2/2}{(1-\alpha^4/4)^{1/2}} \,.
    \label{eq:epsilon true muonium}
\end{align}
Further performing the shift $V_\mu \to V_\mu - \epsilon_\mathrm{eff}{A'}_\mu$ for the ``true muonium'' field and rescaling the dark photon field as $A'_\mu \to A'_\mu/(1-\epsilon_\mathrm{eff}^2)^{1/2}$, we finally remove all kinetic mixing terms from the Lagrangian. 
Then, we obtain the mixing matrix $C_\mathrm{kin}$ for diagonalizing the kinetic terms of $(A', V, A)$ as
\begin{align}
	C_\mathrm{kin} = 	
	\begin{pmatrix}	
		\displaystyle \frac{1}{(1 - \epsilon_\mathrm{eff}^2)^{1/2}} \frac{1}{(1-\epsilon^2)^{1/2}} & 0 & 0 \\
		\displaystyle - \frac{\epsilon_\mathrm{eff}}{(1 - \epsilon_\mathrm{eff}^2)^{1/2}} \frac{1}{(1-\alpha^4/4)^{1/2}} & \displaystyle \frac{1}{(1 - \alpha^4/4)^{1/2}} & 0 \\
		\displaystyle \frac{1}{(1 - \epsilon_\mathrm{eff}^2)^{1/2}} \frac{\epsilon}{(1-\epsilon^2)^{1/2}} + \frac{\epsilon_\mathrm{eff}}{(1 - \epsilon_\mathrm{eff}^2)^{1/2}} \frac{\alpha^2/2}{(1 - \alpha^4/4)^{1/2}} & \displaystyle - \frac{\alpha^2/2}{(1 - \alpha^4/4)^{1/2}} & 1
	\end{pmatrix} \,,
 \label{eq:true muonium mixing matrix}
\end{align}
Combining this and the mass mixing matrix between $V$ and $A'$, we obtain the mixing matrix $C$ relating the interaction basis with the mass basis, as seen in \cref{eq:mass_basis}. 
The mass mixing parameters in the mass matrix (\ref{eq:mass_mixing}) normalized by the mass of the ``true muonium'', $\eta$ and $\delta$, are given as
\begin{align}
    \eta =
    \frac{\epsilon_\mathrm{eff}}{(1-\epsilon^2_\mathrm{eff})^{1/2}},
    \qquad
    \delta =
    \frac{\overline{m}_{A'}/\overline{m}_V}{(1-\epsilon^2_\mathrm{eff})^{1/2}}\,.
    \label{eq:eta true muonium}
\end{align}
Using the mixing matrices $C$ and $C_\mathrm{kin}$ in \cref{eq:true muonium mixing matrix}, the decay widths of the dark photon $A'$, which are calculated using the ``classical'' and the ``mass-insertion'' methods, are obtained as follows:
\begin{align}
    \Gamma_{A'} =
    \frac{\alpha m_{A'}}{3}|C_{AA'}|^2 \, ,
    \qquad 
    \Gamma^\mathrm{MI}_{A'} =
    \frac{\alpha M_{A'}}{3}
    \left|
        (C_\mathrm{kin})_{AA'}
        +(C_\mathrm{kin})_{AV}\frac{-\overline{m}_V^2\eta}{M_{A'}^2-\overline{m}_{V}^2+i\overline{m}_V \overline \Gamma_V}
    \right|^2 \,,
    \label{eq:CMM}
\end{align}
respectively.
Here, $M_{A'}^2 = \overline{m}_{V}^2 (\eta^2 + \delta^2)$ is the mass of the dark photon in the bases with the mass mixing term.
In addition, with $I, J$ representing the vector fields $X, Y$ in the mass basis, i.e., the dark photon and the ``true muonium'', we also obtain the imaginary part of the vacuum polarization, which is required to calculate the decay width in the pole method, as follows:
\begin{align}
	\mathrm{Im}\Pi_{IJ}(s) = -\frac{\alpha}{3} s C_{AI} C_{AJ} \,.
\end{align}

\begin{figure}[t]
    \centering
    \includegraphics[width=0.46\textwidth]{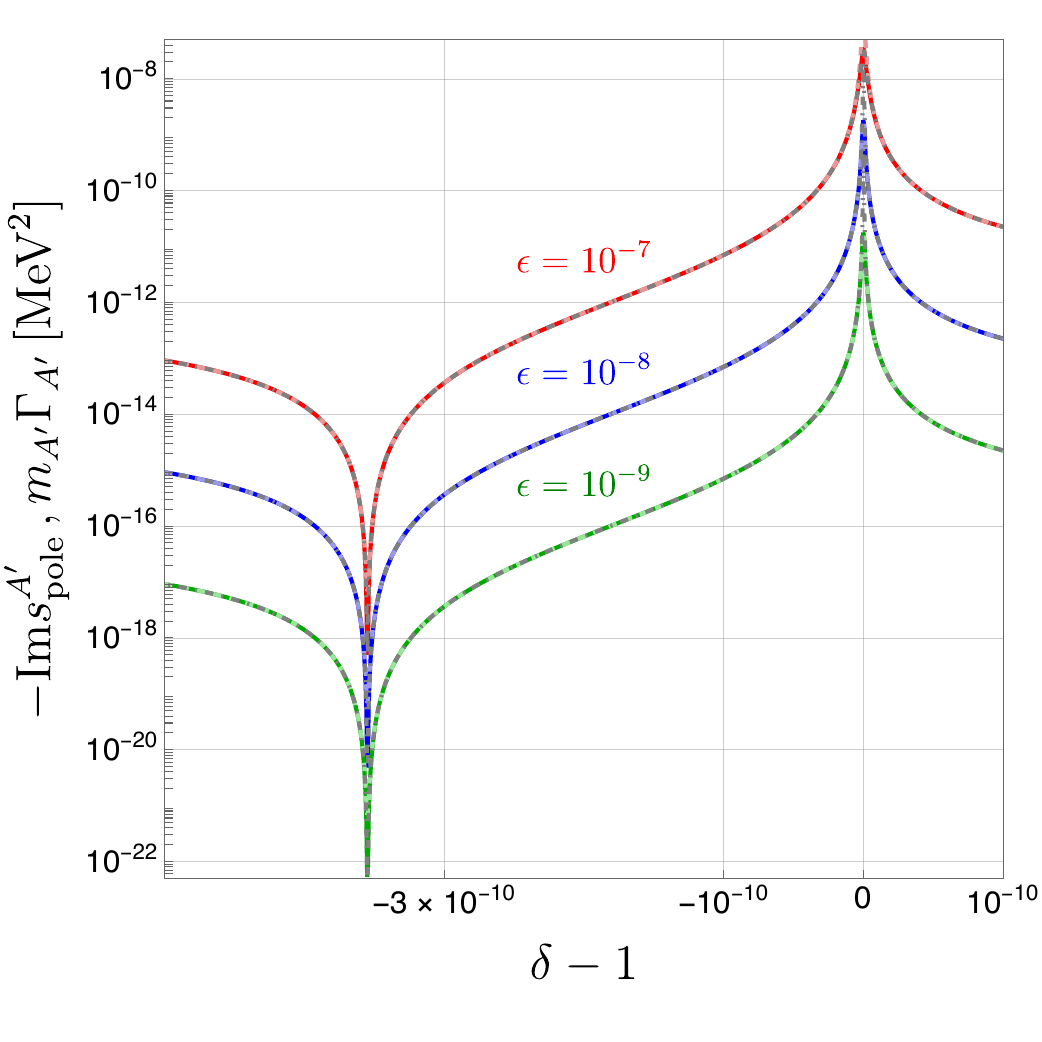}
    \qquad
    \includegraphics[width=0.46\textwidth]{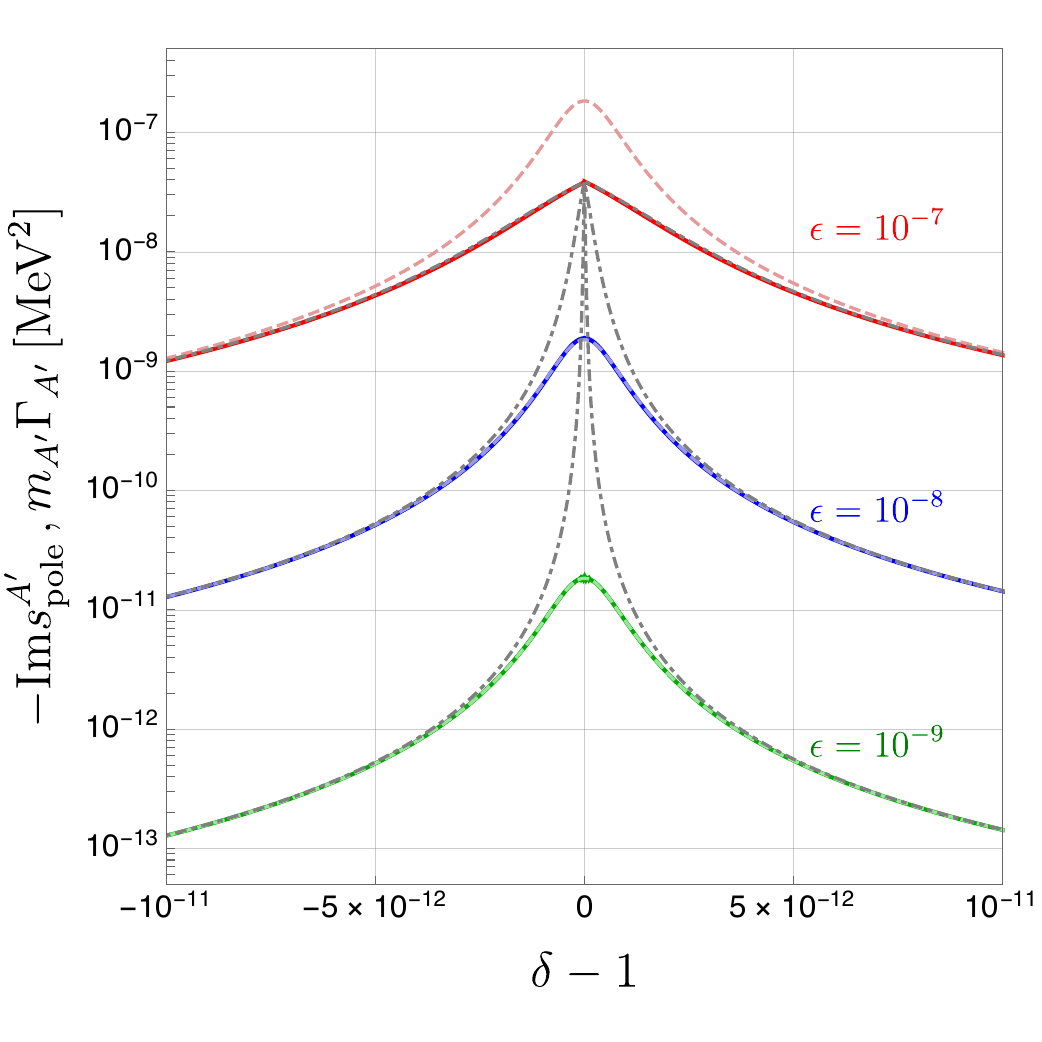}
    \caption{\small\sl
		The same as \cref{fig:Zrate,fig:rhorate}, but for the dark photon $A'$ degenerate with the ``true muonium'' in mass. 
        Each line code is also the same as those in \cref{fig:Zrate,fig:rhorate}. 
        \textbf{Left:} The decay rates as a function of $\delta$ with $\epsilon$ being fixed to be several values. 
        \textbf{Right:} A closer look at the rates with $\delta$ being fixed to be one.
    }
    \label{fig:muoniumrate1}
\end{figure}

\begin{figure}[t]
    \centering
    \includegraphics[width=0.46\textwidth]{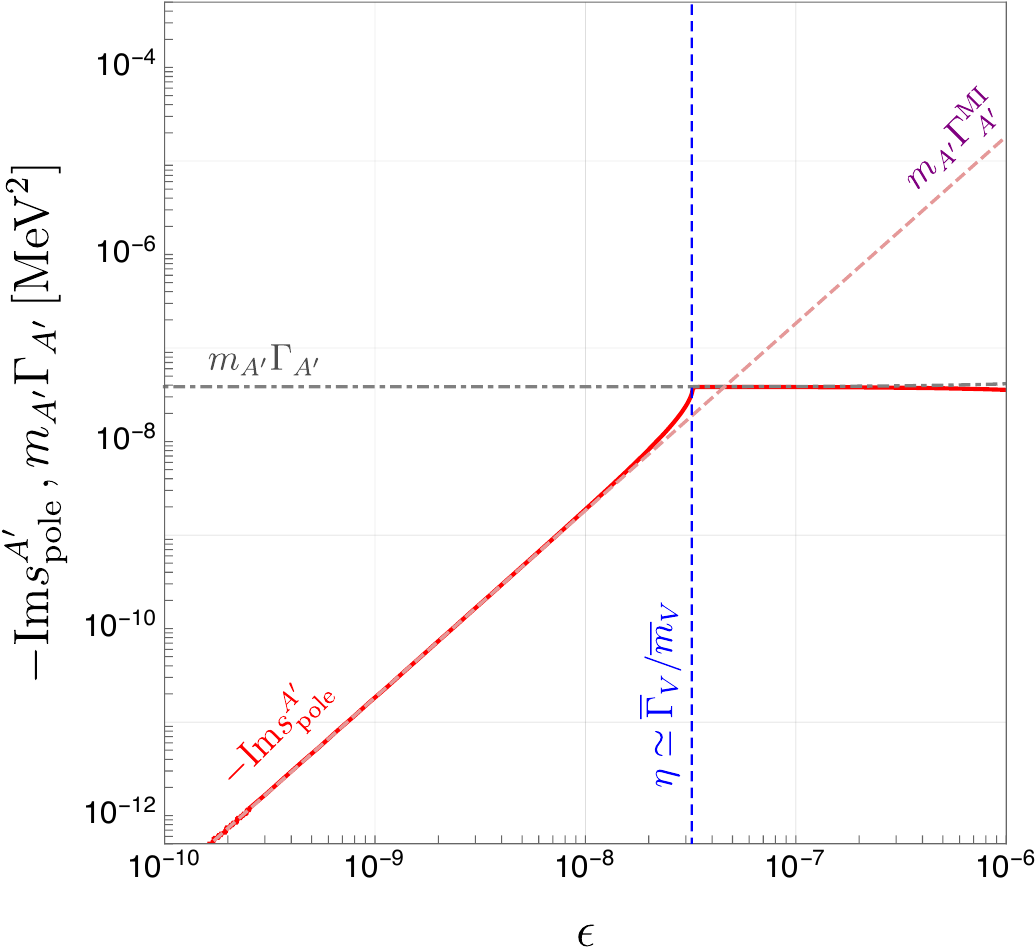}
    \qquad
    \includegraphics[width=0.46\textwidth]{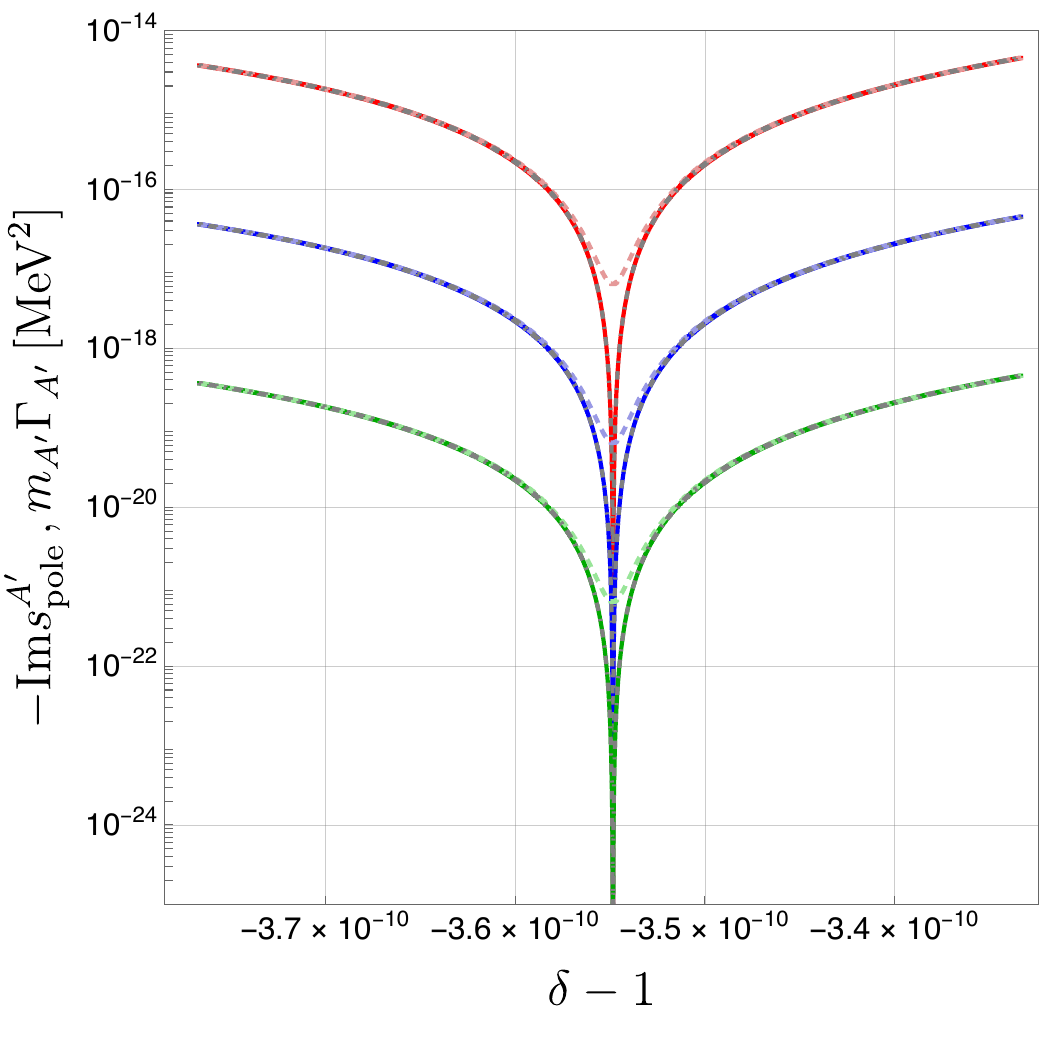}
    \caption{\small\sl
		\textbf{Left:} The same as \cref{fig:Zrate_eps,fig:rhorate_eps}, but for $A'$ degenerate with the ``true muonium'' in mass. 
        Each line code is the same as in \cref{fig:Zrate_eps,fig:rhorate_eps}. 
        \textbf{Right:} A closer look at the rates near the destructive interference.
    }
    \label{fig:muoniumrate2}
\end{figure}

\cref{fig:muoniumrate1} compares the decay widths of the dark photon as a function of $\delta$ with fixed $\epsilon$ ($10^{-7}$, $10^{-8}$, and $10^{-9}$) calculated using the three methods. 
We use the same line types as the previous subsections: colored solid lines for the pole method, colored dashed lines for the ``mass-insertion'' method, and gray dot-dashed lines for the ``classical'' method. 
One can observe that, excluding the region of $\delta \simeq 1$ (the maximum mixing $m_{A'} \simeq m_{V}$), the three methods yield the decay widths in good agreement. 
However, in the vicinity of $\delta = 1$, the ``classical'' method approximates the pole method when $\epsilon = 10^{-7}$, while the ``mass-insertion'' method gives the widths compatible with those of the pole method when $\epsilon$ is smaller, i.e., $\epsilon = 10^{-8}$ and $10^{-9}$. 
This observation is more readily confirmed by examining each width as a function of $\epsilon$ with $\delta$ fixed to be one, as illustrated in the left panel of \cref{fig:muoniumrate2}. 
For the ``true muonium'' case, the kinetic mixing $\epsilon_\mathrm{cr}$ that saturates the criterion $|\eta| \sim \overline \Gamma_V/ \overline m_{V}$ is approximately given by $\epsilon_\mathrm{cr} \simeq 3.2 \times 10^{-8}$ according to \cref{eq:epsilon true muonium,eq:eta true muonium}, and thus the ``classical'' method breaks down as $|\epsilon| > \epsilon_\mathrm{cr}$.

There exists a sharp dip in the decay rates around $\delta \simeq (1 - \alpha^4/4)^{1/2}$ apart from the $|\delta - 1| \lesssim \overline \Gamma_V/\overline{m}_V$ region  as seen in the left panel of \cref{fig:muoniumrate1} (see also the right panel of \cref{fig:muoniumrate2} for a closer look).
In the ``classical'' method, the decay rate $\Gamma_{A'}$ in \cref{eq:CMM} becomes zero when the two contributions to $C_{AA'}$, i.e., the decay directly from the dark photon and that via the ``true muonium'', cancel each other out.
The ``mass-insertion'' method also involves two contributions to $\Gamma^\mathrm{MI}_{A'}$. 
However, the complete cancellation does not occur due to the imaginary part present in the propagator of the ``true muonium'', and the destructive interference is smeared compared to that of the ``classical'' decay width $\Gamma_{A'}$. 
The decay width obtained in the pole method exhibits destructive interference similar to the ``classical'' decay width. 
This fact can be understood by examining the denominator of $D_{A'A'}$ in \cref{eq: propagators}. 
$C_{AA'} = 0$ when the destructive interference occurs.
Since both vacuum polarizations $\Pi_{A'A'}$ and $\Pi_{A'V}$ included in the denominator of $D_{A'A'}$ vanish, the decay width in the pole method gets zero.

\section{Summary and application\label{sec:Dis}} 

Taking the dark photon as an example of a mediator particle connecting the SM and the dark sectors, we introduced the three methods to calculate the decay width of the mediator particle in \cref{sec:Med_Decay}. 
Then, it is found that the decay widths obtained in these methods behave differently when the mediator particle and the SM partner are (nearly-)degenerate in mass in the interaction basis.
\begin{description}
    \item [``Classical'' method] \mbox{} \\
    The decay width of the mediator particle is calculated at tree level in the mass basis, where the field redefinition removes all kinetic and mass mixing among fields relevant to the discussion.
    \item [``Mass-insertion'' method] \mbox{} \\
    We take the basis where the field redefinition removes all kinetic mixing terms while the mass mixing term remains. 
    Regarding the mass mixing between the dark photon and the SM field as a small perturbative parameter, we calculate the decay width by inserting it once.
    \item [Pole method] \mbox{} \\
    Imaginary parts of poles found in current-current correlation functions provide the decay widths of intermediate states, i.e., the mediator particle and its degenerate partner. 
    We obtain the widths by calculating vacuum polarizations in the mass basis and resuming them.
\end{description}
In the ``classical'' method, the dark photon strongly mixes with the SM degenerate partner even if the kinetic mixing parameter $\epsilon$ between their fields is small. As a result, the dark photon has a width comparable to that of the degenerate partner. 
To be more precise, the widths of the two particles become almost half the width the degenerate partner in the absence of $A'$. Note that this fact does not lead to an immediate inconsistency with observations. 
At first glance, halving the width of the degenerate partner also changes the prediction of the production cross-section for its resonance search. 
However, the dark photons are simultaneously produced with the same quantum numbers (and in almost the same amount) as one of the intermediate states of the process, and the total event rate does not change. 
To discriminate the two signals, one from the mediator particle and the other from the degenerate partner, we need the energy resolution to discriminate a tiny mass difference between them. 
Meanwhile, the decay width of the dark photon using the ``mass-insertion'' method is explicitly suppressed by the mass mixing; thus, the dark photon gets long-lived as the kinetic mixing is very small. 
A key feature of the ``mass-insertion'' method is treating the mass mixing perturbatively, as discussed in \cref{sec:Med_Decay}.
However, when the mass mixing is comparable to the decay rate of the degenerate partner, namely $|\eta| \gtrsim \overline \Gamma_V/\overline m_V$, the mass mixing is no longer treated as a perturbative parameter (and it requires the resummation of diagrams with multiple insertions of the mass mixing). 
In such a case, the decay width of the dark photon using the ``classical'' method correctly reproduces the rigorous one, i.e., the decay width using the pole method.

\subsubsection*{Our criterion $|\eta| \simeq \overline \Gamma_V/ \overline m_V$}

In \cref{sec:DP_Decay}, to demonstrate that our criterion for the approximated methods works, in the scenarios with degenerate partners $Z$, $\rho$, and ``true muonium'', we made a numerical comparison among the decay widths of the dark photon using the three methods, assuming that the pole method gives correct decay widths. 
Then, we found the followings: 
(i) when the mass of the dark photon is far away from that of the degenerate partner, the decay widths using the three methods are in good agreement. 
(ii) when the mass difference is comparable to the decay widths of the partner particles, $|\delta-1| \lesssim \overline{\Gamma}_V/\overline{m}_V$, the ``classical'' method and the ``mass-insertion'' method yield different decay rates. 
The ``mass-insertion'' method agrees well with the pole method when $|\eta| \lesssim \overline \Gamma_V/ \overline m_V$, whereas the ``classical'' method does when $|\eta| \gtrsim \overline \Gamma_V/ \overline m_V$. 
This criterion is expected to be universal (as far as low-energy interactions are written in terms of effective field theories); it depends only on the decay rate normalized by its mass and not on other properties of the degenerate partner, e.g., it does not rely on whether the partner is a fundamental particle, a resonance, or a bound state.

\subsubsection*{Application to other resonances}

Let us now consider a general case where the dark photon is degenerate with a vector meson. 
We introduce the critical value of the kinetic mixing parameter between the SM photon and the dark photon, denoted by $\epsilon_\mathrm{cr}$: the ``classical'' (``mass-insertion'') method gives the correct decay rate when the mixing parameter is larger (smaller) than it. 
To determine $\epsilon_\mathrm{cr}$, it is necessary to know the mass mixing parameter $\eta$ between the dark photon and the meson. 
Assuming the vector meson is described by the massive vector field, low-energy interaction terms among the SM photon, the vector meson $H$, and the dark photon are given by
\begin{align}
    \mathcal{L} =
    &
    -\frac{1}{4}F_{\mu \nu}F^{\mu \nu}
    -\frac{1}{4}H_{\mu \nu}H^{\mu \nu}
    -\frac{1}{4}F'_{\mu \nu}F'^{\mu \nu}
    +\frac{\epsilon}{2}F_{\mu \nu}F'^{\mu \nu}
    -\frac{\epsilon_H}{2}F_{\mu \nu} H^{\mu \nu}
    \nonumber \\
    &
    +\frac{1}{2}m_H^2 H_\mu H^\mu
    +\frac{1}{2} m_{A'}^2 A'_{\mu} A'^{\mu}
    +eA_{\mu} J^{\mu}_\mathrm{EM} + \cdots,
\end{align}
where $H_{\mu \nu} = \partial_{\mu} H_{\nu} -\partial_{\nu} H_{\mu}$ and $\widetilde m_H$ are the field strength tensor and the mass of $H$, respectively. 
Here, we assume that (i) the dark photon interacts with the SM through the kinetic mixing between the SM photon and the dark photon, (ii) the meson $H$ interacts with the SM leptons through one photon exchange%
\footnote{
    Multiple photon exchanges are also possible to consider. However, those processes are higher-order corrections to the one photon exchanged diagram in QED. 
    Our prescription used here corresponds to disregarding such corrections.
} 
and its coupling is controlled by the kinetic mixing parameter $\epsilon_H$ between the SM photon and the meson.%
\footnote{
    On the contrary, in addition to the simple photon exchange mentioned above, hadronic currents $J_\mathrm{Had}^\mu$ can couple directly to the vector meson through the current interaction $H_\mu J^\mu_\mathrm{Had}$, as is known in the vector meson dominance.
} 
Following the same discussion as in the previous section, the non-diagonal element of the mass matrix $\eta$ is obtained after removing all the kinetic mixing terms as
\begin{align}
    \eta =
    \frac{\epsilon}{(1 - \epsilon^2)^{1/2}}
    \frac{\epsilon_H}{(1-\epsilon^2_H)^{1/2}} \,.
\end{align}
The vector meson $H$ decays into $e^- e^+$ through the mixing parameter $\epsilon_H$ with the decay rate, 
\eqs{
    \Gamma(H \to e^-e^+) =
    \frac{\overline{m}_H}{16 \pi}
    \frac{4}{3} \epsilon_H^2 e^2 \,,
}
where $\overline{m}_{H}^2 = \widetilde m_{H}^2/(1-\epsilon_H^2)$. 
Using the branching ratio of the vector meson $H$ decaying to a pair of an electron and a positron, $\mathrm{Br}(H \to e^-e^+)$, and the total decay width of the vector meson $\overline{\Gamma}_H$, the unknown parameter $\epsilon_H$ can be written as follows:
\begin{align}
    \epsilon_H =
    \left[\frac{3\,\mathrm{Br}(H \to e^-e^+)\,\overline{\Gamma}_H / \overline m_{H}}{\alpha}\right]^{1/2} \, .
\end{align}
Then, the critical value of the kinetic mixing parameter, $\epsilon_\mathrm{cr}$, which saturates the criterion $|\eta| \simeq \overline{\Gamma}_{H}$ as discussed above, can be quantitatively evaluated in a general case through the following equation:%
\footnote{
    We estimate the critical value $\epsilon_\mathrm{cr}$ using the branching fraction into a $e^- e^+$ pair. 
    Considering the picture of bound states composed of heavy quarkonia and the existence of other decay modes, it may be possible to estimate $\epsilon_\mathrm{cr}$ more accurately. 
    We expect that the order of the magnitude for $\epsilon_\mathrm{cr}$ does not significantly change from our estimate.
}
\begin{align}
    \frac{\epsilon_\mathrm{cr}}{(1-\epsilon^2_\mathrm{cr})^{1/2}}
    \simeq \left[\frac{\alpha\,\overline{\Gamma}_H / \overline m_{H}}{3\,\mathrm{Br}(H \to e^-e^+)}\right]^{1/2} \,.
\label{eq:critical}
\end{align}

\cref{tab:hadrons and critical epsilon} shows various mesons with $J^\mathrm{PC} = 1^{--}$ quantum numbers, their masses, total decay widths, branching ratios into $e^-e^+$, and critical values of the kinetic mixing $\epsilon_\mathrm{cr}$ estimated using \cref{eq:critical}. 
Light vector mesons, i.e., from $\rho\,(770)$ to $\phi\,(1020)$, are composed of light $u\,,d$ and $s$ quarks, whereas heavy vector mesons, i.e., from $J/\psi\,(1S)$ to $\psi\,(4160)$ and from $\Upsilon\,(1S)$ to $\Upsilon\,(11020)$, predominantly made up of $c\bar{c}$ and $b\bar{b}$ quarks, respectively. 
For vector mesons whose dominant decay modes are hadronic and are not suppressed by the Okubo-Zweig-Iizuka (OZI) rule~\cite{Okubo:1963fa,Zweig:1964jf,Iizuka:1966fk}, such as light vector mesons, e.g., $\rho\,(770)$, $\omega\,(782)$, and $\phi\,(1020)$, and heavy quarkonia, e.g., $\psi\,(3770)$ and $\Upsilon\,(4S)$, their total decay widths are large, so that the critical kinetic mixing $\epsilon_\mathrm{cr}$ becomes $\mathcal{O}(1)$. 
Therefore, for the kinetic mixing not excluded by experiments so far searching for the visible decay of the dark photon, the ``mass-insertion'' method provides an accurate prediction of the decay width of the dark photon degenerate with such a vector meson. 
On the other hand, for vector mesons with dominant decay modes suppressed by the OZI rule (e.g., $J/\psi\,(1S)$, $\Upsilon\,(1S)$), $\epsilon_\mathrm{cr}$ is $\mathcal{O}(10^{-3})$ or smaller, so that we have to choose the approximation method correctly, the ``classical'' method and the ``mass-insertion'' method, depending on the magnitude of $\epsilon$.

\begin{table}[t]
    \centering
    \begin{tabular}{r|ccc|c}
        Mesons  & Mass\,(MeV) & Width\,(MeV) & Branching ratio to $e^-e^+$ & Critical mixing $\epsilon_\mathrm{cr}$ \\
       \hline
       $\rho\,(770)$ & $775.26$ & $149.1$ & $4.72 \times 10^{-5}$ & $9.53 \times 10^{-1}$ \\
       $\omega\,(782)$ & $782.66$ & $8.68$ & $7.38 \times 10^{-5}$ & $5.26 \times 10^{-1}$ \\
       $\phi\,(1020)$ & 1019.461 & 4.249 & $2.979 \times 10^{-4}$ & $1.81 \times 10^{-1}$ \\
       $J/\psi\,(1S) $ & 3090.9 & $9.26 \times 10^{-2}$ & $5.971 \times 10^{-2}$ & $1.10 \times 10^{-3}$ \\
       $\psi\,(2S)$ & 3686 & $2.94 \times 10^{-1}$& $7.93 \times 10^{-3}$ & $4.95 \times 10^{-3}$ \\
       $\psi\,(3770)$ &3773.7 &27.2 & $9.6 \times 10^{-6}$ & $8.04 \times 10^{-1}$ \\
       $\psi\,(4040)$ & 4039 & 80 & $1.07 \times 10^{-5}$ & $9.05 \times 10^{-1}$ \\
       $\psi\,(4160)$ & 4191 & 70 & $6.9 \times 10^{-6}$ & $9.25 \times 10^{-1}$ \\
       $\Upsilon\,(1S)$ & 9460 & $5.4\times 10^{-2}$ & $2.38 \times 10^{-2}$ & $7.64 \times 10^{-4}$ \\
       $\Upsilon\,(2S)$ & 10023 & $3.198\times 10^{-2}$ & $1.91 \times 10^{-2}$ & $6.38 \times 10^{-4}$ \\
       $\Upsilon\,(3S)$ & 10355 & $2.032\times 10^{-2}$ & $2.18 \times 10^{-2}$ & $4.68 \times 10^{-4}$ \\
       $\Upsilon\,(4S)$ & 10579.4 & $20.5$ & $1.57 \times 10^{-5}$ & $4.81 \times 10^{-1}$ \\
       $\Upsilon\,(10860)$ & 10885.2 & $37$ & $8.3 \times 10^{-6}$ & $7.67 \times 10^{-1}$ \\
       $\Upsilon\,(11020)$ & 11000 & 24 & $5.4 \times 10^{-6}$ & $7.04 \times 10^{-1}$ \\
       \hline
    \end{tabular}
    \caption{\small\sl 
    Masses, total decay widths, and branching fractions into a $e^- e^+$ pair of vector mesons that could be mixed with the dark photon $A'$, and the critical values of the kinetic mixing \cref{eq:critical}.
    }
    \label{tab:hadrons and critical epsilon}
\end{table}

\section{Concluding remarks \label{sec:Sum}}

We consider the decay of the mediator particle in the presence of a nearly degenerate SM particle by taking the dark photon as an example. 
We find that the ``mass-insertion'' method, in which we treat the mass-mixing parameter (denoted as $\eta$) perturbatively, is valid as far as $\eta$ is smaller than the decay width of the degenerate particle normalized by its mass. 
The ``classical'' method, in which we compute the decay rate in the mass basis, overestimates the decay rate in such a case when the mass difference is smaller than about the decay width of the degenerate partner particle.

\subsubsection*{Implications for existing literature}

The ``classical'' method has been used for $Z$ and $A'$ in the literature.
In the context of collider searches, including electroweak precision tests of the dark photon (or the $Z'$ boson)\,\cite{Hook:2010tw,Curtin:2014cca,Qiu:2023zfr,LoChiatto:2024guj}, and the thermal relic abundance of the inelastic dark matter\,\cite{Izaguirre:2015zva, Izaguirre:2017bqb, Berlin:2018jbm}. 
In the former analysis, the $Z$ boson mass and the couplings of the $Z$ boson to the SM fermions are affected in the presence of the dark photon. 
Then, electroweak observables written in terms of the couplings and the mass are affected. 
When $A'$ is degenerate with $Z$ in mass, it is better to write the observables in terms of the cross-section with re-summed propagators~\cite{Qiu:2023zfr,LoChiatto:2024guj} because the couplings of the $Z$ boson significantly change (with a factor of $1/\sqrt2$). 
Meanwhile, in the latter analysis, the annihilation cross-section of the inelastic dark matter into SM fermions through $A'$ and $Z$ is computed on the mass basis. Considering exclusively $A'$ or $Z$ in the mass basis, the annihilation cross-section would be artificially enhanced when $m_{A'}^2 \simeq m_Z^2$.

\subsubsection*{Use of the $R$-ratio}

The hadronic decay of the dark photon $A'$ is often estimated using a method similar to the ``mass-insertion'' method (e.g., see Refs.\,\cite{Bjorken:2009mm, Blumlein:2013cua, Berlin:2018pwi, Ilten:2018crw}). 
The dark photon couples to the electromagnetic current through kinetic mixing between the dark and SM photons.
Since the theoretical evaluation of the hadron matrix element of the current-current correlation function is not feasible, it is estimated by experimental data, so-called the $R$-ratio, $R(s) = \sigma(e^+ e^- \to \mathrm{hadrons};s)/\sigma(e^+ e^- \to \mu^+ \mu^-;s)$.
More specifically, for the dark photon mass above the hadronic thresholds, the hadronic decay rate is evaluated as
\eqs{
    \Gamma(A' \to \mathrm{hadrons}) =
    R(s=m_{A'}^2)\,\Gamma(A' \to \mu^+ \mu^-) \,.
}
Here $\Gamma(A' \to \mu^+ \mu^-)$ is evaluated in the mass basis for $Z$ and $A'$ (not including hadronic resonances).
The resultant decay rate is always suppressed by a factor of $\epsilon^2$, and thus $\epsilon$ is treated perturbatively as in the ``mass-insertion'' method.
At the same time, the impact on the $R$-ratio on a broad resonance peak is also suppressed by a factor of $\epsilon^2$.
Thus the above seems a good approach near broad resonances even for moderate $\epsilon \sim 0.1$, for instance.
On the other hand, near narrow vector resonances (e.g., $J/\psi\,(1S)$ with $\overline \Gamma_H/ \overline m_H \simeq 3 \times 10^{-5}$, and $\Upsilon\,(1S)$ with $\overline \Gamma_H/ \overline m_H \simeq 6 \times 10^{-6}$), this underestimates the correct decay width of the dark photon $A'$. 
In particular, this would be important for the dark photon prompt decay searches such as BaBar~\cite{BaBar:2009lbr,BaBar:2014zli}, KLOE~\cite{KLOE-2:2011hhj,KLOE-2:2012lii,Anastasi:2015qla,KLOE-2:2016ydq}, and LHCb~\cite{LHCb:2017trq,LHCb:2019vmc}.
In such a case, the ``classical'' method may provide the correct decay rate, which implies that the decay widths of the vector meson and $A'$ are nearly equal to each other (and the half of the vector meson in the absence of $A'$).
As noted above, this does not immediately affect the observables such as $R$-ratio. The mass difference between the vector meson and $A'$ is tiny and thus it may require a high energy resolution to resolve signal events in the experiment into those from the $A'$ decay and those from the decay of the vector resonances.

\subsubsection*{Future work}
This study focuses on the total decay width of the mediator particles. 
Other quantities are also crucial for terrestrial experiments, such as the production cross-section of the mediator particles and the branching fraction into a specific final state. 
We should also consider the influence of the presence of the degenerate partner on these quantities and leave the dedicated analysis of them for a future study. 
We take the dark photon as an example of the mediator particle in \cref{sec:DP_Decay}. 
However, as discussed in \cref{sec:Med_Decay}, the ``mass-insertion'' method is expected to be valid only when the mass mixing is smaller than the decay rate (normalized by the mass) of the nearly degenerate particle $V$ even when we consider other mediator scenarios, such as those with the dark Higgs mediator.

\section*{Acknowledgements}

A. K. acknowledges partial support from Norwegian Financial Mechanism for years 2014-2021, grant nr 2019/34/H/ST2/00707; and from National Science Centre, Poland, grant 2017/26/E/ST2/00135 and DEC-2018/31/B/ST2/02283.
The work of T. K. is supported in part by the National Science Foundation of China under Grant Nos. 11675002, 11635001, 11725520, 12235001, and 12250410248.
S. M. is supported by Grant-in-Aid for Scientific Research from the MEXT, Japan (20H01895, 20H00153, 19H05810, 18H05542, JPJSCCA20200002).
S. M. and Yuki W. are also supported by the World Premier International Research Center Initiative(WPI), MEXT, Japan(Kavli IPMU).
Yu W. is supported by JSPS KAKENHI Grant Number 23KJ0470.

\appendix

\section{SM meson systems \label{app:hadron}}

In this appendix, we discuss that even within the SM, it is possible to find systems similar to the dark photon and its degenerate partner considered in the text and verify our criterion. Specifically, we can find pairs of particles in the hadronic sector that have degenerate masses, consisting of a dark particle in some sense (i.e., similar to the dark photon) and an actively interacting particle (i.e., a degenerate partner) associated with the mixing term due to some symmetry-breaking effect. In such a case, experimental measurements are available in contrast to the scenarios concerning the dark sector discussed in the text. This allows us to clarify which methods are preferable, although both systems involve hadrons, which complicates the discussion. We highlight the $\rho-\omega$ and the neutral $K$ meson systems as concrete examples and comment on our criterion for those systems.

\subsection{\texorpdfstring{$\rho \text{--} \omega \text{--} \gamma$ mixing}{rho--omega--gamma} }\label{app:rhoomegagamma}

There is a pair of nearly degenerate vector particles even in the hadronic resonances without introducing the dark photon, the $\rho$ meson and the $\omega$ meson. Due to the isospin conservation, the $\rho$ meson predominantly decays into two pions, while the $\omega$ meson predominantly decays into three pions.%
\footnote{
This dominant decay channel of $\omega$ meson arises from the gauge invariant 4-form terms, namely, the gauged Wess-Zumino-Witten terms (see for instance, Ref.~\cite{Harada:2003jx}) providing the $\omega$--$\rho$--$\pi$ coupling and $\omega$--$3\pi$ coupling.
}
The three-body state phase space suppresses the $\omega$ width, $\Gamma_\omega \simeq 8.49 \, \mathrm{MeV}$ and makes it much smaller than the $\rho$ width. 
The isospin violation allows us to have the mixing between $\rho$ and $\omega$, and $\omega$ decays into two pions via the mixing.
This implies that this system is analogous to the dark photon and its degenerate partner discussed in the text.

In this subsection, we focus on the isospin-violating decay of the $\omega$ meson, $\omega \to \pi \pi$, and treat $\omega$ as the mediator particle and $\rho$ as its degenerate partner. The mass difference $m_\omega - m_\rho \simeq 7 \, \mathrm{MeV}$ is much smaller than $\overline \Gamma_\rho \simeq 150 \, \mathrm{MeV}$, and hence it is expected that the ``classical'' method and the ``mass-insertion'' method give different decay widths.

It is known that the isospin violation from the electromagnetic charge as introduced in (the $SU(3)$ version of) the fifth term of \cref{eq:LE_Had-DPLag} is not enough for explaining the isospin-violating decay of $\omega$ meson~\cite{OConnell:1995nse}. 
There would be the direct isospin-violating mixing between $\rho$ and $\omega$, $-\epsilon_\mathrm{dir}\rho^0_{\mu\nu}\omega^{\mu\nu}/2$, that originates from the quark mass difference in addition to \cref{eq:LE_Had-DPLag}.
Instead, we give the numerical value of $\epsilon_\mathrm{eff}$ that correctly provides the isospin-violating decay rate of $\omega$ meson in the pole method. 
Using the same $\epsilon_\mathrm{eff}$, we compute the decay rate in the ``classical'' method and the ``mass-insertion'' method and compare them. 
The Lagrangian of this system is 
\eqs{
    \mathcal{L} = - \frac14 F_{\mu\nu} F^{\mu\nu}
    - \frac14 \rho^0_{\mu\nu} \rho^{0 \, \mu\nu}
    - \frac14 \omega_{\mu\nu} \omega^{\mu\nu}
    - \frac{\epsilon_\mathrm{eff}}{2} \rho^0_{\mu\nu} \omega^{\mu\nu}
    + \mathcal{L}_\mathrm{int} \,.
}
Here, the mixing parameter between the $\rho^0$ and $\omega$ mesons is numerically given by  
\eqs{
    \epsilon_\mathrm{eff}
    \simeq 0.0078 \,, \quad 
    \eta = \frac{\epsilon_\mathrm{eff}}{\sqrt{1 - \epsilon_\mathrm{eff}^2}} \simeq 0.0078 \,.
}
Using $\overline m_\omega = 782.65 \, \mathrm{MeV}$, and $\overline m_\rho = 775.26 \, \mathrm{MeV}$, we find the numerical value of the mass mixing matrix parameters as
\eqs{
    \delta = \frac{\overline{m}_{\omega} / \overline{m}_\rho}{\sqrt{1 - \epsilon_\mathrm{eff}^2}} \simeq 1.01\,.
}
Since the numerical $\eta$ is quite smaller than $\overline \Gamma_\rho/\overline m_\rho \simeq 0.2$, the ``mass-insertion'' method gives an approximate width of $\omega$ meson into two pions. 
We consider the basis where the kinetic terms are diagonalized and treat the mass mixing between $\rho^0$ and $\omega$ as a perturbative parameter as discussed in the text.
The mixing matrix for diagonalizing (and canonically normalizing) the kinetic terms $C_\mathrm{kin}$ is given by
\eqs{
	C_\mathrm{kin} = 
	\begin{pmatrix}
		\displaystyle \frac{1}{(1 - \epsilon_\mathrm{eff}^2)^{1/2}} \frac{1}{(1-(3 e/g)^2)^{1/2}} & 0 & 0 \\
		\displaystyle - \frac{\epsilon_\mathrm{eff}}{(1 - \epsilon_\mathrm{eff}^2)^{1/2}} \frac{1}{(1-e^2/g^2)^{1/2}} & \displaystyle \frac{1}{(1 - e^2/g^2)^{1/2}} & 0 \\
		\displaystyle \frac{1}{(1 - \epsilon_\mathrm{eff}^2)^{1/2}} \frac{3 e/g}{(1-(3 e/g)^2)^{1/2}} + \frac{\epsilon_\mathrm{eff}}{(1 - \epsilon_\mathrm{eff}^2)^{1/2}} \frac{e/g}{(1 - e^2/g^2)^{1/2}} & \displaystyle - \frac{e/g}{(1 - e^2/g^2)^{1/2}} & 1
	\end{pmatrix} \,.
}
Here, $e/g$ denotes the kinetic mixing between the $\rho$ meson and photon, and $3 e/g$ denotes the kinetic mixing between the $\omega$ meson and photon.
The relation to the effective kinetic mixing is given by 
\eqs{
    \epsilon_\mathrm{eff} = \epsilon_\mathrm{dir} - \frac{e/g}{\sqrt{1 - e^2/g^2}} \frac{3 e/g}{\sqrt{1-(3 e/g)^2}} \,.
}
Here, $\epsilon_\mathrm{dir}$ denotes the direct isospin-violating kinetic mixing and would originate from the quark mass difference. 
In the ``mass-insertion'' method, the decay rate of $\omega$ meson into two pions (with assuming massless final states) is 
\eqs{
	\Gamma^\mathrm{MI}_{\omega}
	& = \frac{M_{\omega}}{16 \pi} \frac13 \left| 
	\overline g_{\pi}^{\omega} + \overline g_{\pi}^{\rho} \frac{- \eta \overline m_\rho^2}{M_{\omega}^2-\overline m_\rho^2+i \overline m_\rho \overline \Gamma_\rho} \right|^2 \,, \\
	\overline g_{\pi}^{\omega} & \equiv (C_\mathrm{kin})_{\rho \omega}\,g + (C_\mathrm{kin})_{A\omega}\,e \,, \quad 
	\overline g_{\pi}^{\rho} \equiv (C_\mathrm{kin})_{\rho \rho}\,g + (C_\mathrm{kin})_{A\rho}\,e \,.
}
Here, $M_\omega^2 =\overline m_\rho^2 (\eta^2 + \delta^2)$. 
We obtain its numerical value as $\Gamma^\mathrm{MI}_{\omega} \simeq 0.22 \,\mathrm{MeV}$ using $\overline \Gamma_\rho$ given by \cref{eq:barGammarho}, the $\rho\pi\pi$ coupling $g = 5.92$, and the electromagnetic fine structure constant $\alpha^{-1}_\mathrm{EM} = 137$. 
On the other hand, in the ``classical'' method, the decay rate is 
\eqs{
    \Gamma_{\omega}
	& = \frac{m_{\omega}}{16 \pi} \frac13 (g_{\pi}^{\omega})^2 \,, \qquad 
	g_{\pi}^{\omega} \equiv C_{\rho \omega}\,g + C_{A\omega}\,e  \,.
}
Its numerical value is $\Gamma_{\omega} \simeq 20 \,\mathrm{MeV}$. 
Finally, the imaginary part of the vacuum polarization is computed as 
\eqs{
    \mathrm{Im} \Pi_{IJ} (s)
	& = - \frac{\pi s}{16 \pi^2} \frac{1}{3} g^I_{\pi} g^J_{\pi} \,,
}
where $I \,, J$ run $\rho$ and $\omega$. 
The imaginary part of the pole corresponding to the $\omega$ meson is numerically $- \mathrm{Im}(s_\omega)/\sqrt{\mathrm{Re}(s_\omega)} \simeq 0.13 \, \mathrm{MeV}$.
Meanwhile, from the central value of the PDG values of the total decay width of $\omega$ meson $\Gamma_\omega \simeq 8.49 \, \mathrm{MeV}$ and the branching fraction into two pions $\mathrm{BR}(\omega \to \pi \pi) \simeq 1.53 \, \%$~\cite{ParticleDataGroup:2022pth}, we obtain $\Gamma(\omega \to \pi \pi) \simeq 0.130 \,\mathrm{MeV}$. 

This system has $|\delta - 1| \leq \overline \Gamma_\rho/\overline m_\rho \simeq 0.2$, and $|\eta| \ll \overline \Gamma_\rho/\overline m_\rho \simeq 0.2$. 
As expected from the $\rho$-$A'$ system in the text (see the right panel of \cref{fig:rhorate}), the ``mass-insertion'' method provides the partial decay width that approximates the width in the pole method, while the ``classical'' method provides the partial decay width much larger than the width in the pole method.
We note that the decay width with the ``mass-insertion'' method does not precisely agree with the width in the pole method. 
This is because the strong coupling $g$ leads to sizable corrections to the real part of poles from the imaginary parts of the vacuum polarization.

\subsection{Kaon mixing \label{app:kaon}}

Our criterion is applicable to other systems with other spins even though we consider a vector mediator particle as an example so far.
In this subsection, we comment on the neutral $K$ meson system as an application of three methods to spin-zero particles. 
The neutral kaons $K^0$ and $\overline K^0$ are exactly degenerate in mass as far as the weak interactions are turned off, i.e., within QED and QCD. 
Once turning on the weak interactions (without $CP$ violation), the mass basis denoted by ($K_1$, $K_2$) is no longer the eigenstate of the strangeness due to the flavor-changing process. Note that this basis is still eigenstate of the $CP$ transformation, and $K_1$ and $K_2$ denote the $CP$ even and odd eigenstates, respectively. Due to the $CP$ conservation, $K_1$ decays to two pions, while $K_2$ decays to three pions, and hence it leads to make the $K_2$ width smaller than the $K_1$ width.
When the $CP$ violation is turned on, the $CP$ eigenstates ($K_1$, $K_2$) are mixed with each other, and the mass eigenstates are denoted by $K_L$ and $K_S$. 
$K_S$ is predominantly composed of $K_1$ and primarily decays into two pions, on the other hand $K_L$ is predominantly composed of $K_2$ and primarily decays into three pions.
$K_L$ also decays into two pions through the $CP$ violation. 

From the above discussion, we can consider the neutral $K$ meson system as an scalar version of our analysis. Taking $(K^0, \overline{K}^0)$ as Schr\"odinger field variables which describe the non-relativistic neutral $K$ mesons, we effectively write their Heisenberg equation as
\eqs{
    i \frac{d}{dt} 
    \begin{pmatrix}
        K^0 \\ 
        \overline{K}^0
    \end{pmatrix} 
    = \left( M - \frac{i}{2} \Gamma \right) 
    \begin{pmatrix}
        K^0 \\ 
        \overline{K}^0
    \end{pmatrix} \,,
    \label{eq:Heisenberg}
}
where both $M$ and $\Gamma$ are Hermitian matrices constituting the effective Hamiltonian. 
The effective Hamiltonian consists of the strong interaction and the weak interaction: the diagonal components are predominantly determined by the strong interactions, while the off-diagonal components and the difference of the diagonal components are from the weak interactions in the $(K^0, \overline K^0)$ basis. 
After partially diagonalizing the $CP$-symmetric part, we may regard the $(K_1, K_2)$ basis as the basis with the mass matrix similar to \cref{eq:massmat}. 
The mixing parameter $\eta$ corresponds to the $CP$-violating part of the (weak) effective Hamiltonian.
The eigenvalues of the effective Hamiltonian determine the pole positions of the propagators, allowing us to identify the poles as $K_S$ with the large imaginary part and $K_L$ with the small imaginary part, respectively.

The experimental observations tell us the information about the mass mixing matrix.  
The known mass difference between $K_L$ and $K_S$ gives $\Delta M_K = M_{K_L} - M_{K_S} \simeq - 2 \mathrm{Re} M_{12} \simeq 3.5 \times 10^{-12} \, \mathrm{MeV}$, and the decay rate of $K_S$ (denoted by $\gamma_S$) is $\gamma_S \simeq 7.4 \times 10^{-12} \, \mathrm{MeV}$, corresponding to the lifetime of $\tau_{K_S} \simeq 0.9 \times 10^{-10} \, \mathrm{s}$. 
The mixing of fields is known as the indirect $CP$ violation and represents the difference of the mass basis ($K_S, K_L$) and the $CP$ basis ($K_{1}, K_{2}$),
\eqs{
    K_S \simeq K_1 - \epsilon K_2 \,, \quad
    K_L \simeq K_2 - \epsilon K_1 \,.
}
The indirect $CP$ violation is parametrized in the complex mass parameters, and its absolute value is measured to be 
\eqs{
    |\epsilon| \simeq \frac{1}{2 \sqrt{2}} \frac{\mathrm{Im}(M_{12})}{\mathrm{Re}(M_{12})} \simeq 2.2 \times 10^{-3} \,.
}
The mass difference between $K_{1,2}$ is much larger than the $CP$-violating mass mixing. 
This implies that the size of the mass mixing parameters $\eta$ and $\delta$ so that $\eta \ll |1-\delta|$.
From our analysis, it is expected that all methods give the decay rates at the same orders of amplitude when the mass difference normalized by its mass is larger than $\eta$ and the mass difference is comparable to the width of the degenerate particle ($\Delta M_K/\gamma_S \simeq 0.5$): namely $\delta - 1 \simeq \Delta M_K/M_K$ is close to the dimensionless decay width $\gamma_S/M_K$. 
This parameter set corresponds to the near-boundary where the three methods begin to deviate from each other in the right panels of \cref{fig:Zrate,fig:rhorate,fig:muoniumrate1}.

In the literature, we find a seemingly different method in contrast to our method discussed in this subsection for calculating the decay width of the neutral $K$ meson, known as the Wigner-Weisskopf method. There, the two-state wave-function for $K^0$ and $\overline{K}^0$ is introduced, and its time evolution is described by the effective Schr\"{o}dinger equation with the Hamiltonian similar to the one in \cref{eq:Heisenberg}. The decay widths are then identified as the imaginary part of the eigenvalues of the Hamiltonian. However, this method is essentially the same as what we refer to as the pole method, and it is merely a translation of the second-quantized formulation into the language of first quantization.

There are several caveats for the application to the meson system. 
In our examples in the text, the dark photon does not have a direct coupling to the SM particles in the basis where the kinetic mixing between the dark photon and the degenerate particle remains. 
However, in the kaon system, there is the $\Delta S = 1$ process (so called direct $CP$ violation). We need to account for the direct coupling in addition to the couplings appeared after removing kinetic and mass mixings. 
The mass difference $\Delta M_K$ is comparable with the width of $K_S$, $\gamma_S$, in the kaon system. 
Since the ``classical'' method does only incorporate the mass difference and the mass mixing, the ``classical'' method would provide the $CP$-violating decay width of $K_L$ slightly different from the width using other two methods due to the sizable width $\gamma_S$.

\section{Example: \texorpdfstring{$\rho$}{rho} dominantly decaying into electrons\label{app:rhoee}}

We discuss again the dark photon decaying through the mixing with the $\rho$ meson discussed in \cref{sec:DP_Decay}. 
In the SM, the dominant decay mode of the $\rho$ meson is that into a $\pi^+ \pi^-$ final state. 
In this appendix, we consider a hypothetical case where the $\rho$ meson dominantly decays into a $e^+ e^-$ pair; even so, we assume the same coupling $g = 5.92$ and the same mass $m_\rho = 775.26$\,MeV. 
The electrons couple to the vector fields only through the electromagnetic current.
Hence, the $\rho$ meson decays into a pair of electrons only through the kinetic mixing between the $\rho$ meson and photon, and the decay rate of the $\rho$ meson at the tree level is suppressed by the kinetic mixing,
\eqs{
	\overline \Gamma_{\rho}^{ee}
	= \frac{m_{\rho}}{16 \pi} \frac43 \left|\frac{e/g}{\sqrt{1-e^2/g^2}} e \right|^2 \,, 
}
and numerically $\overline \Gamma_{\rho}^{ee} = 4.96$\,keV. 
This decay channel provides an example of the resonance with $\overline \Gamma_{\rho}^{ee}/\overline m_\rho \simeq 6 \times 10^{-6}$, which is between two $\overline \Gamma_V/ \overline m_V$s for the pion channel of the $\rho$ meson ($\overline{\Gamma}_\rho/\overline{m}_\rho \simeq 0.2$) and for the ``true muonium'' ($\overline{\Gamma}_V/\overline{m}_V \simeq 10^{-12}$). 
The decay rate of the dark photon calculated in the ``classical'' method is given by
\eqs{
	\Gamma_{A'}^{ee}
	= \frac{m_{A'}}{16 \pi} \frac43 |C_{AA'} e|^2 \,.
}
Here, the mixing matrix $C_{AA'}$ is given in \cref{eq:Cmix_rho,eq:mass_basis}. 
Meanwhile, the decay rate by the ``mass-insertion'' method is
\eqs{
    \Gamma^{ee; \mathrm{MI}}_{A'} & =
    \frac{M_{A'}}{16 \pi} \frac43
    \left| 
        \overline g_{e}^{A'}
        +\overline g_{e}^{\rho}
        \frac{-\eta \overline m_\rho^2}{M_{A'}^2 - \overline m_\rho^2 + i \overline m_\rho \overline \Gamma_{\rho e}}
    \right|^2 \,,
    \\
    \overline g_{e}^{A'} & \equiv
    (C_\mathrm{kin})_{AA'}\,e \,,
    \quad 
    \overline g_{e}^{\rho} \equiv (C_\mathrm{kin})_{A\rho}\,e \,.
    \label{eq:rhoee_MI}
}
Since the tree-level coupling $\overline g_{e}^{A'}$ does not vanish in the electron channel in contrast to the pion channel, there appears a cancellation between two contributions from the direct coupling and from the mass insertion for a certain parameters. 
Finally, the imaginary part of the vacuum polarization from the electron loop is given by
\eqs{
    \mathrm{Im} \Pi_{IJ} (s) & =
    -\frac{\pi s}{16 \pi^2} \frac{4}{3}\,g_{e}^{I}\,g_{e}^{J} \,.
}
Here, $g_{e}^{I}$ in the above formula denotes the coupling of the electron to the vector field $I$.

\begin{figure}[t]
    \centering
    \includegraphics[width=0.46\textwidth]{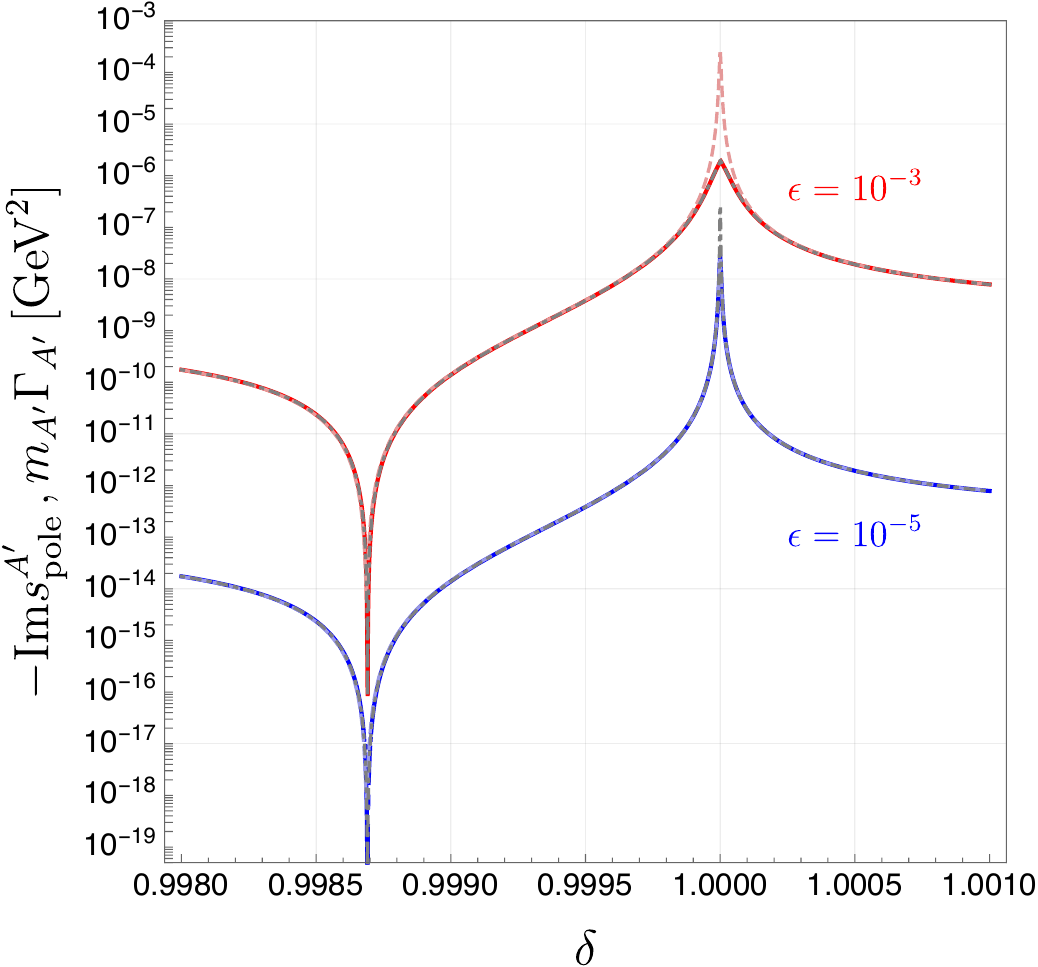}
    \qquad
    \includegraphics[width=0.46\textwidth]{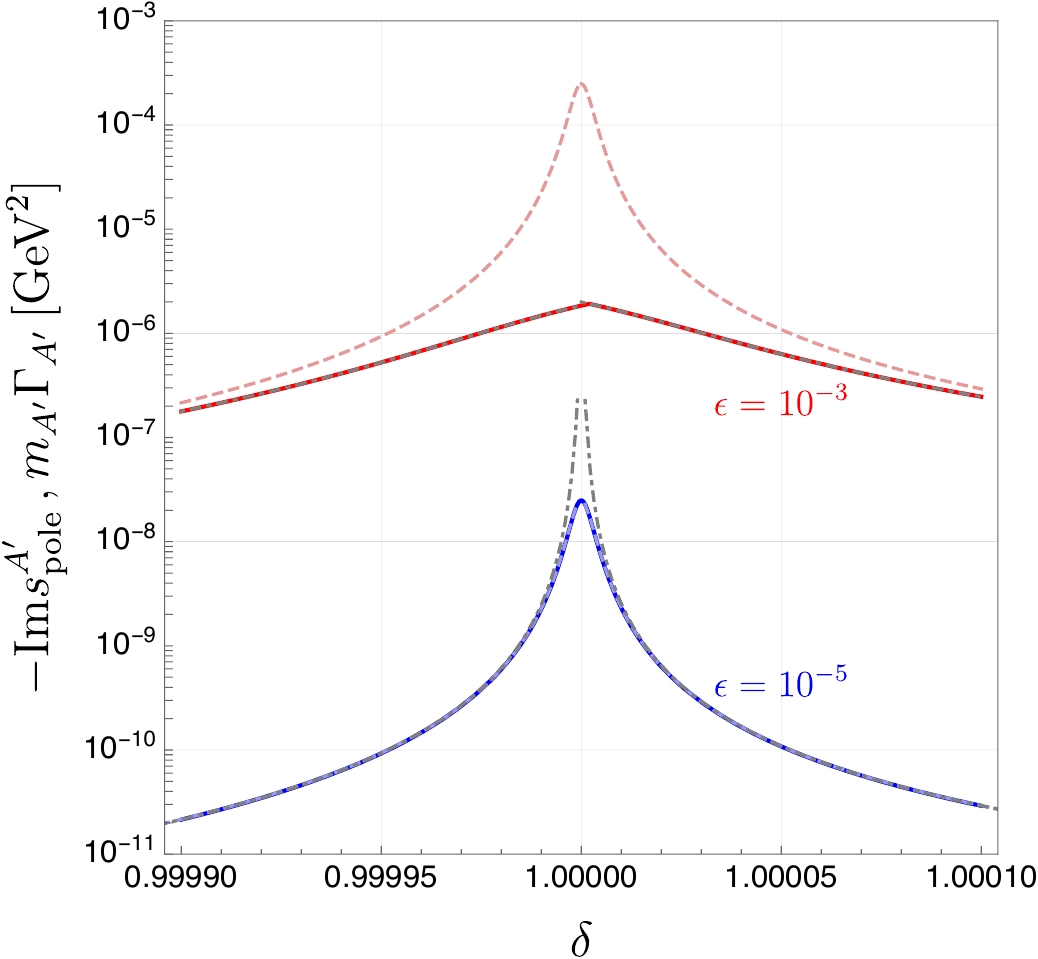}
    \caption{\small \sl
		The same as \cref{fig:rhorate}, but $A'$ decays only into $e^+ e^-$. Each line code is the same as that in \cref{fig:rhorate}. \textbf{Left:} The decay rates as functions of $\delta$ for a fixed $\epsilon$.
		\textbf{Right:} A closer look at decay rates near $\delta = 1$.
    }
    \label{fig:rhoeerate}
\end{figure}

We show in \cref{fig:rhoeerate} the comparison of the total decay rates calculated using the three methods. 
We choose two different values of $\epsilon$: $\epsilon = 10^{-3}$ (red) and $\epsilon = 10^{-5}$ (blue). 
As expected, the decay rates using the three methods are in agreement with each other except for the region with $\delta \simeq 1$. 
When $\delta$ approaches to unity (within the range of $\overline \Gamma_{\rho}^{ee}/ \overline m_\rho \simeq 6 \times 10^{-6}$), the three methods are not compatible with each other. 
The ``mass-insertion'' method (depicted as colored-dashed lines) gives the width in agreement with the width by the pole method when $|\eta| \lesssim \overline \Gamma_{\rho}^{ee}/ \overline m_\rho$ (or $|\epsilon| \lesssim 10^{-4}$). 
On the other hand, the ``classical'' method (depicted as dot-dashed lines) gives the width in agreement with the width by the pole method when $|\eta| \gtrsim \overline \Gamma_{\rho}^{ee}/ \overline m_\rho$ (or $|\epsilon| \gtrsim 10^{-4}$). 
We also see a destructive interference near $\delta \simeq 0.9987$ in the right panel of \cref{fig:rhoeerate}. 
Let us consider the decay width using the ``mass-insertion'' method in \cref{eq:rhoee_MI}. 
Since the location of the destructive interference is enough away from $\delta = 1$, we can ignore the imaginary part of the propagator, $i \overline m_\rho \overline \Gamma_{\rho}^{ee}$. 
Then, we find the decay rate vanishes at $\delta \simeq (1 - e^2/g^2)^{1/2} \simeq 0.9987$.

\begin{figure}[t]
    \centering
    \includegraphics[width=0.5\textwidth]{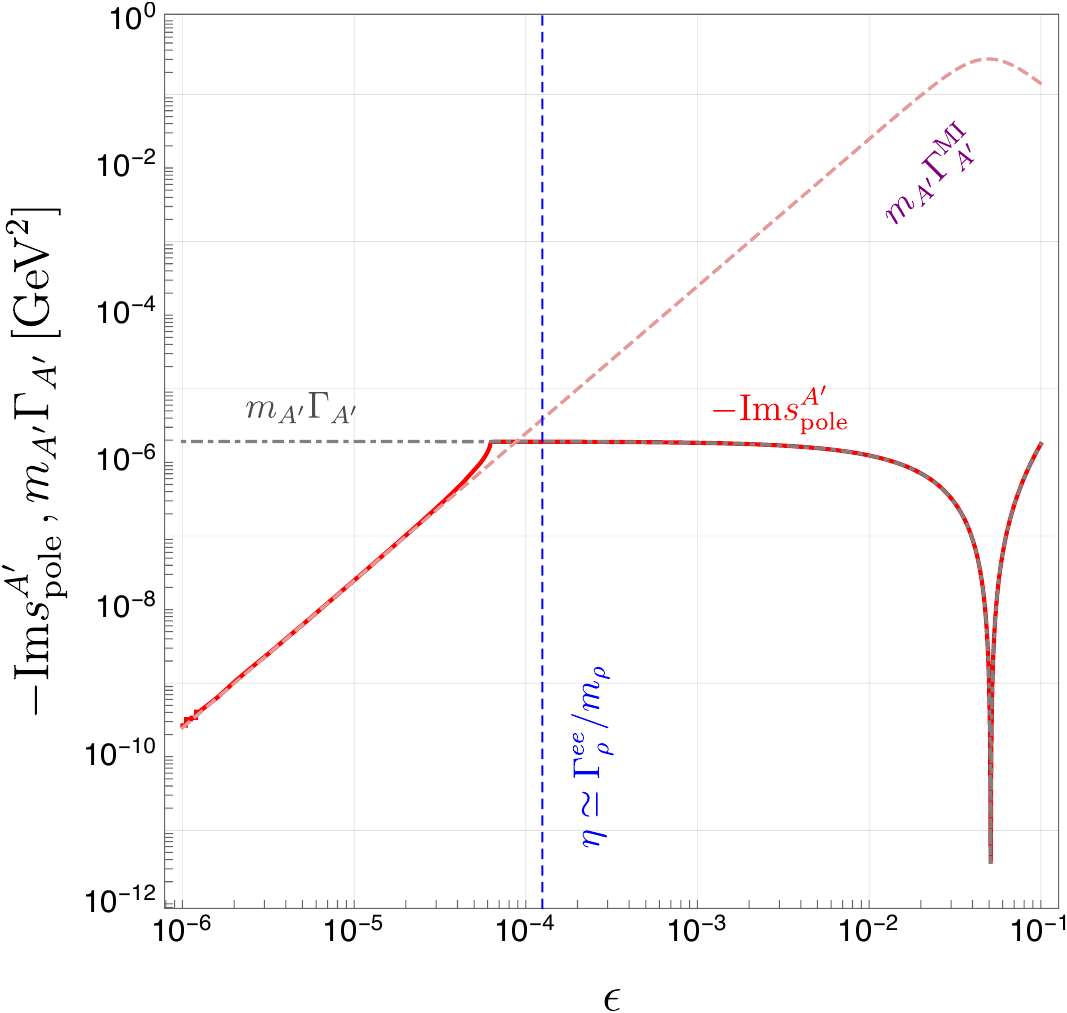}
    \caption{\small \sl
		The same as \cref{fig:rhorate}, but $A'$ decays only into $e^+ e^-$. Each line code is the same as in \cref{fig:rhorate}.
    }
    \label{fig:rhoeerate_eps}
\end{figure}

As in \cref{sec:DP_Decay}, we also show the comparison of the rates as a function of $\epsilon$ with $\delta = 1$ in \cref{fig:rhoeerate_eps}. 
When $\epsilon$ is larger than the threshold\,(i.e., given by $|\eta| \simeq \overline \Gamma_{\rho}^{ee}/ \overline m_\rho$), the decay rate of $A'$ by the imaginary part of the pole agrees with that by the ``classical'' method. 
This behavior is consistent with the conclusion in the text. A destructive behavior is again seen at $\epsilon \simeq 0.05$. 
There, the coupling of the dark photon to an electron and a positron in the mass basis,
\eqs{
    C_{AA'} =
    \frac{e/g}{\sqrt{1-e^2/g^2}} \cos \theta
    +\eta
    \left(
        \frac{e/g}{\sqrt{1-e^2/g^2}}
        -\frac{\sqrt{1-e^2/g^2}}{e/g}
    \right) \sin\theta  \,,
}
vanishes at a specific $\eta$ for $\delta = 1$, where the mass mixing is almost maximum, $\theta \simeq + \pi/4$. 
Then, we find $\eta \simeq e^2/g^2$ (or $\epsilon \simeq e/g \simeq 0.05$) by solving $C_{AA'} = 0$. 
We note that, as discussed in \cref{sec:Med_Decay}, the particles are identified by their decay rates near $\delta = 1$. 
The dark photon $A'$ in mass basis corresponds to the mass eigenstate whose eigenvalue is given by $m_Y^2$ in \cref{eq:mass_mixing} at $\delta$ far from unity.
On the other hand, near $\delta = 1$, the dark photon $A'$ corresponds to the mass eigenstate whose eigenvalue is given by $m_X^2$.
Therefore, at $\delta$ far from unity, we should use the coupling, 
\eqs{
	C_{AA'} = - \frac{e/g}{\sqrt{1-e^2/g^2}} \sin \theta + \eta \left( \frac{e/g}{\sqrt{1-e^2/g^2}} + \frac{\sqrt{1-e^2/g^2}}{e/g} \right) \cos\theta \,.
}.

\section{Kinetic mixing with true muonium\label{app:muonium_detail}}

We show the detail of determining the mixing parameter $\kappa$ between the SM photon and the (ground-state) ``true muonium'' in the effective field theory (EFT)\,(\ref{eq:eta true muonium}) and deriving the decay width of the ``true muonium''. 
To obtain $\kappa$, we match the amplitude of the $V \to e^- e^+$ process in QED and the EFT. 
First, in the QED calculation, we assume that the state of the ``true muonium'', i.e.,a non-relativistic bound state, can be written by a superposition of a pair of the freely moving non-relativistic $\mu^-$ and $\mu^+$ states as follows:
\begin{align}
    \ket{V,S,S_z} \simeq
    \sqrt{2 m_V}
    \sum_{s_1,s_2} \sigma^{SS_z}_{s_1 s_2}
    \int \frac{d^3p}{(2\pi)^3}\,\tilde{\psi}(\vec{p})
    \frac{1}{\sqrt{2m_\mu}} \ket{\mu^-; +\vec{p}, s_1}
    \otimes
    \frac{1}{\sqrt{2m_\mu}} \ket{\mu^+; -\vec{p}, s_2} \,,
    \label{eq:state of true muonium}
\end{align}
where $\sigma^{SS_z}_{s_1 s_2}$ is the wave function of the spin describing the spin-triplet state, while $\tilde{\psi}(\vec{p})$ is the Fourier transformation of the wave function $\psi(\vec{r})$ that describes the muon-antimuon system in the position space. 
Note that $\psi(\vec{r})$ is obtained by solving the Schr\"{o}dinger equation with the Hamiltonian $H = -\nabla^2/m_\mu - \alpha/r$. 
Moreover, $\ket{\mu^-\,(\mu^+); \vec{p}, s}$ is the state vector of the muon\,(antimuon) with the momentum $\vec{p}$ and the spin $s$. 
Here, to normalize the state in a covariant way, we put a factor $1/\sqrt{2 m_\mu}$. 
Using the expression of the ``true muonium'' in \cref{eq:state of true muonium}, we obtain the amplitude as
\begin{align}
    \mathcal{M}^\mathrm{QED} (V \to e^-e^+) =
    \frac{1}{\sqrt{m_\mu}}
    \sum_{s_1,s_2} \sigma^{SS_z}_{s_1s_2}
    \int \frac{d^3p}{(2\pi)^3}\,\tilde{\psi}(\vec{p})
    \,\mathcal{M}^\mathrm{QED}
    [(\mu^-; +\vec{p}, s_1)\,(\mu^+; -\vec{p}, s_2)
    \to e^- e^+] \,,
    \label{eq:QED amplitude}
\end{align}
where $\mathcal{M}^\mathrm{QED}[(\mu^-; +\vec{p}, s_1)\,(\mu^+; -\vec{p}, s_2) \to e^- e^+]$ is the amplitude from a pair of $\mu^-,\,\mu^+$ into $e^-\,e^+$ with muon-antimuon momenta and spins being $\vec{p},\,-\vec{p}$ and $s_1\,,s_2$, respectively.
This amplitude is explicitly given by
\begin{align}
    \mathcal{M}^\mathrm{QED}[(\mu^-; +\vec{p}, s_1)\,(\mu^+; -\vec{p}, s_2) \to e^- e^+] =
    \frac{e^2}{m_V^2}
    [\bar{v}_{s_2}(-\vec{p})\gamma_\mu u_{s_1}(\vec{p})]
    \,
    [\bar{u}(\vec{k})\gamma^\mu v(-\vec{k})] \,,
\end{align}
with $u(\vec{p})$ and $v(\vec{p})$ being the spinor wave functions associated with the external lines, and $\vec{k}$ being the momentum of the outgoing electron $e^-$. Remembering $\mu^-$ and $\mu^+$ are non-relativistic, $u_s(\vec{p})$ and $v_s(-\vec{p})$ becomes independent of their momenta, and the wave function part in \cref{eq:QED amplitude} is factored out, and its value is simply given by the wave function at the origin in the position space, i.e., $\psi(0) = (\alpha^3 m_\mu^3/8\pi)^{1/2}$.
On the other hand, in the EFT, we obtain the amplitude as follows:
\begin{align}
    \mathcal{M}^\mathrm{EFT}(V \to e^-e^+) =
    e\,\kappa\,\epsilon_\mu (S_{z})\,
    \bar{u}(\vec{k})\gamma^\mu v(-\vec{k}) \,,
    \label{eq:EFT amplitude}
\end{align}
where $\epsilon_\mu$ is the polarization vector of the ``true muonium''. From the matching condition on \cref{eq:QED amplitude,eq:EFT amplitude}, we find $\kappa = \alpha^2/2$.%
\footnote{
    The spinor wave functions are $u_s(\vec{p}) \simeq \sqrt{2m_\mu}(\chi_s,0)^\mathrm{T}$ and $v_s(-\vec{p}) \simeq -\sqrt{2m_\mu}(0,\chi_s')^\mathrm{T}$ in the Dirac representation with $\chi_s$ and $\chi_s'=-i\sigma^2\chi_s^*$ being the two-component spinor, and $\chi_{s=+1}=(1,0)^\mathrm{T}$ and $\chi_{s=-1}=(0,1)^\mathrm{T}$. The polarization vectors are $\epsilon^\mu (S_{z}=1) = -(0,1,i,0)/\sqrt{2}$, $\epsilon^\mu (S_{z}=0) = (0,0,0,1)$ and $\epsilon^\mu (S_{z}=-1) = (0,1,-i,0)/\sqrt{2}$.
}  In addition, following the standard procedure from amplitude to a physical quantity, \cref{eq:QED amplitude} gives the decay width of the ``true muonium'' as $\Gamma_V = \alpha^5 m_\mu/6$.

\bibliographystyle{utphys}
\bibliography{ref}

\end{document}